\documentclass[aps,prd,nofootinbib,twocolumn,superscriptaddress,preprintnumbers,balancelastpage,longbibliography]{revtex4-1}
\usepackage{aas_macros}

\usepackage{placeins}
\usepackage{amsmath,amssymb,mathtools,bm}
\usepackage{graphicx, color, hepunits}
\usepackage[dvipsnames]{xcolor}
\usepackage{float}
\usepackage{filecontents}
\usepackage{wasysym}

\usepackage{multirow}
 \usepackage{hyperref} 
\hypersetup{
    colorlinks=true,       
    linkcolor=blue,        
    citecolor=blue,        
    filecolor=magenta,     
    urlcolor=blue          
}

\usepackage{slashed}

\usepackage{url}
\usepackage[utf8]{inputenc}
\usepackage[english]{babel}
\usepackage{mathrsfs}

\newlength{\dhatheight}

\newcommand{\es}[2] {\begin{equation} \label{#1} \begin{split} #2 \end{split} \end{equation}}

\usepackage{placeins}

\begin{document}

\title{
Wino and Real Minimal Dark Matter Excluded by Fermi Gamma-Ray Observations
}

\preprint{CERN-TH-2025-137}

\author{Benjamin R. Safdi}
\email{brsafdi@berkeley.edu}
\affiliation{Berkeley Center for Theoretical Physics, University of California, Berkeley, CA 94720, U.S.A.}
\affiliation{Theoretical Physics Group, Lawrence Berkeley National Laboratory, Berkeley, CA 94720, U.S.A.}
\affiliation{Theoretical Physics Department, CERN, 1211 Geneva, Switzerland}

\author{Weishuang Linda Xu}
\email{wlxu@slac.stanford.edu}
\affiliation{Kavli Institute for Particle Astrophysics \& Cosmology, Stanford University, Stanford, CA 94305, U.S.A.}
\affiliation{Particle Theory Group, SLAC National Accelerator Laboratory, Stanford, CA 94305, U.S.A.}

\date{\today}

\begin{abstract} 
We show that minimal, fermionic dark matter (DM) models in the $n$-dimensional representation of the weak force with zero hypercharge that make up 100\% of the DM under the standard cosmological history are strongly excluded for $n < 9$.  This includes the thermal wino, which we show is ruled out even allowing for DM core sizes up to $\sim$6.7 kpc with the preferred local DM density or $\sim$3.7 kpc in addition to the local DM density being half the preferred value, at less than $0.2$ GeV/cm$^3$.  We reach these conclusions through dedicated searches with 14 years of Fermi gamma-ray data in the inner Galaxy between 30 GeV and 2 TeV for the continuum gamma-rays produced in the decays of unstable particles produced in DM annihilation and bound-state formation processes.  We consider a variety of Milky Way DM profiles in our analyses, including those motivated by modern hydrodynamic cosmological simulations, and show that all the $n < 9$ minimal DM models are disfavored even under the most conservative assumptions for these density profiles. 
While wino, quintuplet ($n=5$), and $n = 7$  DM models are strongly disfavored by our analyses under the standard cosmology, we discuss how non-standard cosmological histories or DM sub-fractions could still allow for these particles to be realized in nature, with discovery opportunities at next-generation particle colliders and gamma-ray telescopes.    
\end{abstract} 

\maketitle

\section{Introduction}

Minimal dark matter (DM)~\cite{Cirelli:2005uq} is the simplest realization of the weakly interacting massive particle (WIMP) DM paradigm. In this framework the DM is embedded in a non-trivial representation of the electroweak force with hypercharge assignment such that there is a neutral component.
Below the scale of electroweak symmetry breaking the charged components of the DM electroweak multiplets receive loop corrections to their masses that make them heavier than the neutral components, which are viable DM candidates.  The two simplest fermionic minimal DM candidates are the higgsino (electroweak doublet) and the wino (electroweak triplet). These particles are especially well motivated because they naturally emerge in the context of supersymmetric (SUSY) extensions of the Standard Model (SM), where they may also mix with the bino (electroweak singlet), see~\cite{Jungman:1995df,Lisanti:2016jxe,Slatyer:2017sev,Hooper:2018kfv,Safdi:2022xkm,Cirelli:2024ssz} for reviews.  
In particular, these DM candidates arise in split-spectrum SUSY scenarios such as `mini-split' SUSY and variants~\cite{Wells:2003tf,Giudice:2004tc,Arkani-Hamed:2004ymt,Hall:2011jd,Arvanitaki:2012ps,Arkani-Hamed:2012fhg}, where the scalar superpartners are parametrically heavier than the fermionic superpartners. Split-spectrum SUSY models sacrifice fine tuning for the electroweak hierarchy problem to maintain the accuracy of precision grand unification and consistency with the Higgs boson mass, while also avoiding otherwise troublesome flavor changing neutral currents found in low-scale SUSY.  
As we show, however, wino DM is strongly excluded by indirect searches for DM annihilation even for the most conservative assumptions about the DM profile of the Milky Way.  

To achieve the correct relic abundance through the WIMP freeze-out mechanism, assuming the universe is radiation dominated from freeze-out to big bang nucleosynthesis (BBN), the wino should have a mass $m_\chi \approx 2.8$ TeV.  We refer to the wino with this mass as the thermal wino.
It has been known for over a decade that the thermal wino is in tension with searches for gamma-rays produced by DM annihilation in our galaxy~\cite{Fan:2013faa,Cohen:2013ama,Rinchiuso:2018ajn,Rodd:2024qsi}. 
Refs.~\cite{Fan:2013faa,Cohen:2013ama} in particular used gamma-ray data from the ground-based High Energy Stereoscopic System (HESS) to conclude that the thermal wino is excluded unless the Milky Way DM profile has a $\sim$kpc core.  On the other hand, as we discuss in this work, a kpc-scale core would not be surprising for a Milky-Way-like galaxy.  

Refs.~\cite{Fan:2013faa,Cohen:2013ama} searched for the gamma-ray line at the end-point of the annihilation spectrum, which arises primarily from the one-loop-level $\gamma\gamma$ final state that is significantly enhanced by the Sommerfeld effect at the low velocities found in our Galaxy.  In this work, in contrast, we use Fermi Large Area Telescope (LAT) gamma-ray data to search for the continuum gamma-ray emission at lower energies, in particular below $\sim$2 TeV, induced primarily from the showering of the tree-level $W^+ W^-$ final state.  We show that the Fermi continuum search has superior sensitivity relative to the HESS line search, especially for more cored DM profiles, since the Fermi analysis is able to incorporate data from further away from the Galactic Center.  
We rule out the wino even allowing for the possibility of $\sim$6 kpc DM cores, which are already far larger than simulations support for galaxies like the Milky Way~\cite{Butsky:2015pya,2020MNRAS.497.2393L}. We thus conclude that the thermal wino is not realized in nature.  

The search we perform with Fermi-LAT data for the wino is similar to that performed to search for evidence for continuum emission from higgsino DM annihilation~\cite{Dessert:2022evk}. That analysis found mild evidence in favor of a higgsino signal, whose thermal mass is $\sim 1.1$ TeV, though in this analysis we find no evidence in favor of the thermal wino.  We also show that searches for DM annihilation in Milky Way dwarf galaxies~\cite{Fermi-LAT:2010cni,Fermi-LAT:2011vow,Fermi-LAT:2013sme,Fermi-LAT:2015att,Fermi-LAT:2015ycq,Geringer-Sameth:2014qqa,Fermi-LAT:2016uux,Calore:2018sdx,Hoof:2018hyn,McDaniel:2023bju}
 independently rule out the thermal wino DM paradigm.  If $m_\chi$ is less than 2.8 TeV, then the wino is only a subfraction of the DM, unless the cosmological history is modified. In the subfraction scenario we are able to robustly rule out all wino masses between 1.85 TeV and the thermal mass, with the wino being less than 50\% of the DM at the lower end.

In contrast to indirect detection, the wino is notoriously difficult to probe with DM direct detection experiments, with estimates putting the expected scattering cross-section near the neutrino floor~\cite{Hill:2011be,Hill:2013hoa,Hill:2014yka,Hill:2014yxa,Hisano:2015rsa}. Collider probes of winos with the ATLAS and CMS detectors have ruled out the minimal model up to masses $\sim$700 GeV by searching for disappearing tracks~\cite{CMS:2020atg,ATLAS:2022rme}, independent of cosmological abundance.  The wino is a benchmark target for future colliders such as a future 10 TeV muon collider or 100 TeV proton-proton collider such as the Future Circular Collider-hh (FCC-hh)~\cite{Han:2020uak,Capdevilla:2021fmj,Cirelli:2014dsa}; in both cases, achieving 5$\sigma$ discovery with winos with masses at or below $\sim$4 TeV could be achievable. On the other hand, as we show in this work, the thermal wino as a DM candidate is strongly excluded. Collider probes could still be useful, however, in cases where the wino is a DM subfraction or where {\it e.g.} a period of early matter domination necessitates that the wino have a larger mass to explain the DM abundance. 

In addition to wino DM we also consider quintuplet ($5$-plet) and higher-dimensional odd, real representations of $SU(2)_L$ for fermionic minimal DM.  
In the context of minimal DM, the Lagrangian of the SM, ${\mathcal L}_{\rm SM}$, is extended to~\cite{Cirelli:2005uq}
\es{eq:MDM}{
{\mathcal L}_{\rm MDM} = {\mathcal L}_{\rm SM} + c \bar \chi\left(  i \slashed{D} + m_\chi \right) \chi\,,
}
where $c = 1$ ($c = 1/2$) for a Dirac (Majorana) fermion $\chi$.  The DM candidate $\chi$ must be a singlet under the strong force, and we take it to be in the $n$-dimensional representation of $SU(2)_L$, referred to as the $n$-plet. Recall that the electric charge is given by $Q = T_3 + Y$, with $T_3$ the third generator of $SU(2)_L$ in the given representation and $Y$ the hypercharge of $\chi$.  The $n$-plet representation of $SU(2)_L$ has $T_3 = {1 \over 2}{\rm diag}(n-1, n-3, \cdots, -n+1)$.  The simplest minimal DM models, giving an electrically neutral component to $\chi$, are thus those with odd $n$ and $Y = 0$, which are real representations. The content of the these multiplets $\chi$ is a tower of $n$ states whose electric charges span the range from $-(n-1)/2$ to $+(n-1)/2$. The neutral Majorana component, labeled $\chi^0$, is the lightest and thus a suitable DM candidate, with chargino counterparts lifted in mass by radiative corrections, {\it e.g.} $m_\chi^+ - m_\chi \approx 164$ MeV at 2-loop order~\cite{Ibe:2012sx}, after electroweak symmetry breaking.  

DM candidates within these representations
are almost entirely unprobed experimentally, in part because they do not scatter off of nuclei through $Z$-boson exchange,  automatically evading direct detection constraints~\cite{Bottaro:2021snn,Bloch:2024suj}.  While the wino ($n = 3$) is the simplest and canonical example of minimal, real, fermionic DM with $Y = 0$, the quintuplet ($n = 5$) is also strongly motivated~\cite{Cirelli:2005uq,Baumgart:2023pwn}; in particular, the quintuplet is cosmologically stable without necessitating additional symmetries, and the electroweak theory does not introduce low-scale Landau poles. It is additionally interesting that quintuplet and higher electroweak representations are not known to arise in isolation in string theory constructions, and thus their discovery may have profound implications on the nature of our UV theory~\cite{Baumgart:2024ezp}.

Given the Lagrangian~\eqref{eq:MDM} the cosmological history of the dark sector $\chi$ is fully determined, at least assuming a standard radiation dominated cosmological history from temperatures $T \sim M_\chi$ down to BBN. (Periods of early matter domination would modify the cosmological history in a straightforward way, allowing for larger DM masses.)  Additionally, we assume that the lightest component of $\chi$ is stable on cosmological time-scales, though depending on the representation of $\chi$ this may require additional symmetries~\cite{Cirelli:2005uq}. 
In Tab.~\ref{tab:MDM} we summarize the DM masses $m_\chi$ giving the correct abundance of DM for each representation of $SU(2)_L \times U(1)_Y$ that we consider in this work. These masses were were computed at leading order (LO) in Refs.~\cite{Mahbubani:2020knq,Bottaro:2022one}, with the exception of the wino, whose thermal mass was computed at next-to-leading-order (NLO) in Ref.~\cite{Beneke:2020vff}. The sources of theory uncertainty in these numbers is discussed in  Ref.~\cite{Bottaro:2023wjv}, which become large for the heavier candidates in large $n$ representations, dominated by NLO corrections 
as the electroweak coupling approaches perturbative unitarity.  

One promising way to search for these minimal electroweak DM candidates is to target the sharp line-like endpoint of the gamma-ray annihilation spectrum, coming from processes emitting at least one photon with energy close to the DM mass. As these thermal candidates typically have masses in the $\mathcal{O}(10-100)$ TeV range, this signal is best searched for with ground-based pointing telescopes with sensitivity to ultra-high-energy gamma rays, such as the upcoming Cherenkov Telescope Array Observatory (CTAO). Discovery prospects for several of these minimal scenarios at these observatories have been studied in {\it e.g.}~\cite{Cohen:2013ama,Rinchiuso:2020skh,Abe:2025lci,Baumgart:2023pwn,Cirelli:2015bda}. On the other hand, the annihilation of these heavy DM particles into unstable electroweak gauge bosons still induces a large soft ``continuum" spectrum that reaches well into the sensitivity window of Fermi, and this is the signal that we target here.

We show in this work that, in addition to the wino, the quintuplet and 7-plet are strongly disfavored by the dedicated searches with Fermi-LAT data that we perform. The 9-plet, on the other hand, is still allowed, even if less theoretically appealing. 
Higher dimensional representations are not constrained by our analysis but run into issues with {\it e.g.} perturbative unitarity~\cite{Bottaro:2022one}. 
Crucial to our conclusions are the theoretical calculations, which we perform building off of {\it e.g.}~\cite{Baumgart:2023pwn}, of the continuum annihilation cross-sections accounting for Sommerfeld enhancement~\cite{Hisano:2003ec,Hisano:2004ds} and bound-state formation and decay processes. Our calculations of these rates are summarized in the next section, with Sec.~\ref{sec:data} describing our Fermi data analysis procedure, Sec.~\ref{sec:wino} summarizing our results for the wino, and Sec.~\ref{sec:nb5} summarizing our results for the $n>3$ minimal DM models.

\begin{table}[t]
\begin{tabular}{|c|c|c|c|c|c|c|}
\hline
$n_Y$ & $m_\chi$ [TeV]    & $\mu_{\rm Thelma}^{\rm 95}$ & $\mu_{\rm Einasto}^{\rm 95}$  & $\rm{sign}(\hat \mu) \times {\rm TS}_{\rm Thelma}$   \\ \hline
$ 3_0 $               & $2.86 \pm 0.01$ &  0.12 & 0.07 & -0.79 \\ \hline
$ 5_0  $            & $13.6 \pm 0.8$ & 0.41 & 0.29 & -0.73 \\ \hline
$7_{0}$ & $48.8 \pm 3.3$ & 0.8 & 0.55 & -0.60 \\ \hline
$9_{0}$ & $113 \pm  15$ & 2.1 & 1.5  & -0.52
\\ \hline
\end{tabular}
\caption{\label{tab:MDM} 
Summary of the upper limits on the minimal DM annihilation cross-sections from the analyses performed in this work of Fermi continuum gamma-ray data above 30 GeV towards the Galactic Center. We label the fermionic, real minimal DM models by their quantum numbers $n_Y$, with $n$ the dimension of the $SU(2)_L$ representation and $Y=0$ representing zero hypercharge. The mass column $m_\chi$ shows the masses, and associated uncertainties, where the minimal DM models may explain 100\% of the DM assuming radiation-dominated cosmology prior to BBN. The upper limits on the annihilation cross-section are displayed through the $\mu$ parameters, which rescale the expected minimal DM annihilation cross-sections with $\mu=1$ representing the expected cross-section. Thus, upper limits below unity imply that the model is ruled out by our analysis. We provide the weakest upper limit scanning over the range of possible masses in the mass column assuming respectively the Einasto profile and the significantly more conservative ``Thelma" profile, see text for details. Additionally, for the Thelma profile analysis we provide the discovery TS for the two-sided test in favor of the minimal DM model times the sign of the best-fit annihilation cross-section. All of our fiducial analyses have slightly negative best-fit annihilation cross-sections, though at less than 1$\sigma$ significance.  We conclude that all but the $9_0$ model are excluded.   }
\end{table}

\section{Annihilation of minimal DM}
\label{sec:min}

In all the minimal DM models the DM candidate $\chi^0$ is a Majorana fermion, such that the differential flux of gamma-rays from DM annihilation in an angular direction $\Omega$ is ({\it e.g.}, see~\cite{Safdi:2022xkm})
\es{eq:ann_formula}{
{d N \over dE d\Omega} = {J(\Omega) \,{\mathcal E}(E,\Omega) \over 8 \pi m_{\rm DM}^2} {d \langle \sigma v \rangle_\gamma \over dE} \,,
}
with units of cts/GeV/sr. Here, \mbox{$J \equiv  \int_0^\infty ds \, \rho_{\rm DM}^2(s,\Omega)$} is the $J$-factor that is an integral over the DM density squared along the line of sight in the angular direction specified by $\Omega$. The energy- and direction-dependent exposure of the instrument is specified by ${\mathcal E}(\Omega)$, which has units of cm$^2$ s.  The differential annihilation cross-section for continuum emission to photons may be written as
\es{}{
{d \langle \sigma v \rangle_\gamma \over dE} \approx {\langle \sigma v \rangle_{WW} {dN_{WW} \over dE}}+{\langle \sigma v \rangle_{ZZ + \frac{1}{2} \gamma Z} {dN_{ZZ} \over dE}} \,,
}
where for the moment we ignore emission from bound-state formation and decay and where we also ignore higher-order corrections to the end-point of the spectrum, including the monochromatic contribution from  $\gamma \gamma$ and $\gamma Z$ processes, since we are primarily interested in the low-energy continuum. (We return to the spectral endpoint later in this section, as it is relevant for cross-checks we perform on our wino results.)  Note that ${d \langle \sigma v \rangle_\gamma \over dE}$ is the differential annihilation spectrum normalized {\it per annihilation}; that is, ${\langle \sigma v \rangle_{WW}}$ (${\langle \sigma v \rangle_{ZZ}}$) is the annihilation cross-section to $W^+W^-$ ($ZZ$), while ${dN_{WW} \over dE}$ (${dN_{ZZ} \over dE}$) is the differential number of photons produced per annihilation to $WW$ ($ZZ$).  We compute the differential spectra using \texttt{PPPC4DMID}~\cite{Buch:2015iya} (for lower DM masses) and \texttt{HDMSpectra}~\cite{Bauer:2020jay} (for $m_\chi \gtrsim$ 100 TeV). 

The tree-level annihilation cross-sections ${\langle \sigma v \rangle_{WW}}$ and ${\langle \sigma v \rangle_{ZZ}}$ are fully specified by the representation of $\chi$ under $SU(2)_L \times U(1)_Y$ and the DM velocity distribution, with little dependence on the comparatively tiny charged mass splittings between the states in $\chi$. However, these cross-sections, particularly in the present day where DM is highly non-relativistic, are strongly affected by Sommerfeld enhancement.

\subsection{Sommerfeld enhancement}

The annihilation rates of these minimal DM candidates are highly sensitive to a number of electroweak effects, increasingly so as the particle mass and representation dimensionality under consideration increases.  The foremost of these effects is Sommerfeld enhancement: the exchange of virtual electroweak bosons, functionally long-range for the scales of $\gtrsim$ TeV DM annihilation, creates a substantial attractive potential that hugely modifies the incoming wavefunctions. This effect can be computed by numerically solving the multi-state time-independent Schr\"{o}dinger equation 
\begin{equation}
    \psi_i'' (x)  = \left( \frac{V^{ij} (x)}{E} + \frac{L (L+1)}{x^2} -1 \right) \psi_j (x), 
\label{eq:sf}\end{equation}
where $\psi_i$ represents the reduced wavefunction of the annihilating 2-particle state, and it is convenient to work in dimensionless radial measure $x \equiv p r$.  The momentum $p = m_\chi v/2$ and kinetic energy $E = p^2/m_\chi$ are defined relative to the neutral DM 2-body state, where $v$ is the halo velocity. These velocities are very small, $v^2\sim 10^{-6}$, so the kinetic energy carried by the $\mathcal{O}(1-100 \text{ TeV})$ DM candidates we consider are generally insufficient to produce on-shell chargino pairs. Thus far away from the interaction potential the two-particle state is well-described by pure  $\chi^0\chi^0$. 

Since total charge ($Q$), spin ($S$), and angular momentum ($L$) are conserved quantities of the system, the wavefunction $\psi_i = \left\{ \chi^0\chi^0, \chi^+ \chi^-,  \dots , \chi^{(n-1)/2} {\bar \chi}^{(n-1)/2} \right\}$ spans the  $(n+1)/2$ allowed 2-body states, where for brevity we define  $\chi^i \equiv \chi^{\overbrace{+\dots +}^{i}}$, ${\bar \chi}^i \equiv \chi^{\overbrace{- \dots -}^{i}} $ to refer to the charginos with $i$ units of electric charge. The boundary condition on this equation is such that $\psi_i( x\to \infty) \sim \left\{ {\rm e}^{ikx}, 0, \dots, 0 \right\}$. The $s$-wave annihilation is obtained upon solving~\eqref{eq:sf} with $L=0$.  

The electroweak potential $V^{ij}$ can be decomposed into several pieces: a term accounting for the mass gaps between various particle states, and terms corresponding to various electroweak gauge boson exchange processes,    
\begin{equation}
        V^{ij}(x) = V_\delta^{ij} + V_{W}^{ij}(x) + V_{Z\gamma}^{ij}(x). 
\label{eq:Vew_q0}\end{equation}
Explicitly, 
\begin{align}
    V_\delta^{ij} & = \frac{2 i^2 \delta m \, m_\chi  }{p^2} \delta^{ij}\\
    V_W^{ij}(x) &= -  \frac{C_{W,i}^2 \alpha_W m_\chi}{x p } \left( {\rm e}^{- m_W x/p} + \zeta^W_{\rm NLO} (x) \right) \nonumber\\ &  \times \left(\delta^{i, j+1} + \delta^{i+1, j} \right)\\
    V_{Z\gamma}^{ij}(x) &= -  \frac{ i^2 \alpha_W m_\chi  }{x p } \left( s_W^2 + c_W^2 {\rm e}^{- m_Z x/p} + \zeta^Z_{\rm NLO} (x) \right) \delta^{ij}, 
\end{align}
where $\delta  m= 164$ MeV is the basic mass gap set by radiative corrections (with $m_{\chi^i} - m_{\chi^0} = i^2 \delta m$ ), $s_W (c_W) = \sin\theta_W (\cos \theta_W)$,  and 
\begin{equation}
    C_{W, i} g_W = \frac{n+2i+1}{2\sqrt{2}} g_W \sqrt{\frac{(\frac{n+1}{2}-i)! (\frac{n+1}{2}+i)!}{(\frac{n-1}{2}-i)! (\frac{n+3}{2}+i)!} }
\end{equation} is the coupling associated with the $\chi^i {\bar \chi}^{i+1} W$ vertex, obtained by expanding~\eqref{eq:MDM} in the appropriate representation. The terms $\zeta^{W,Z}_{\rm NLO}$ represent corrections to the electroweak potential at NLO; we use the analytic fitting functions presented in~\cite{Beneke:2019qaa,Urban:2021cdu}. As these corrections modify the precise locations of the Sommerfeld resonances, the predictions for the Galactic annihilation cross-section of {\it e.g.} the thermal mass could be substantially modified. 

\begin{figure}[!htb]
\includegraphics[width = 0.45\textwidth]{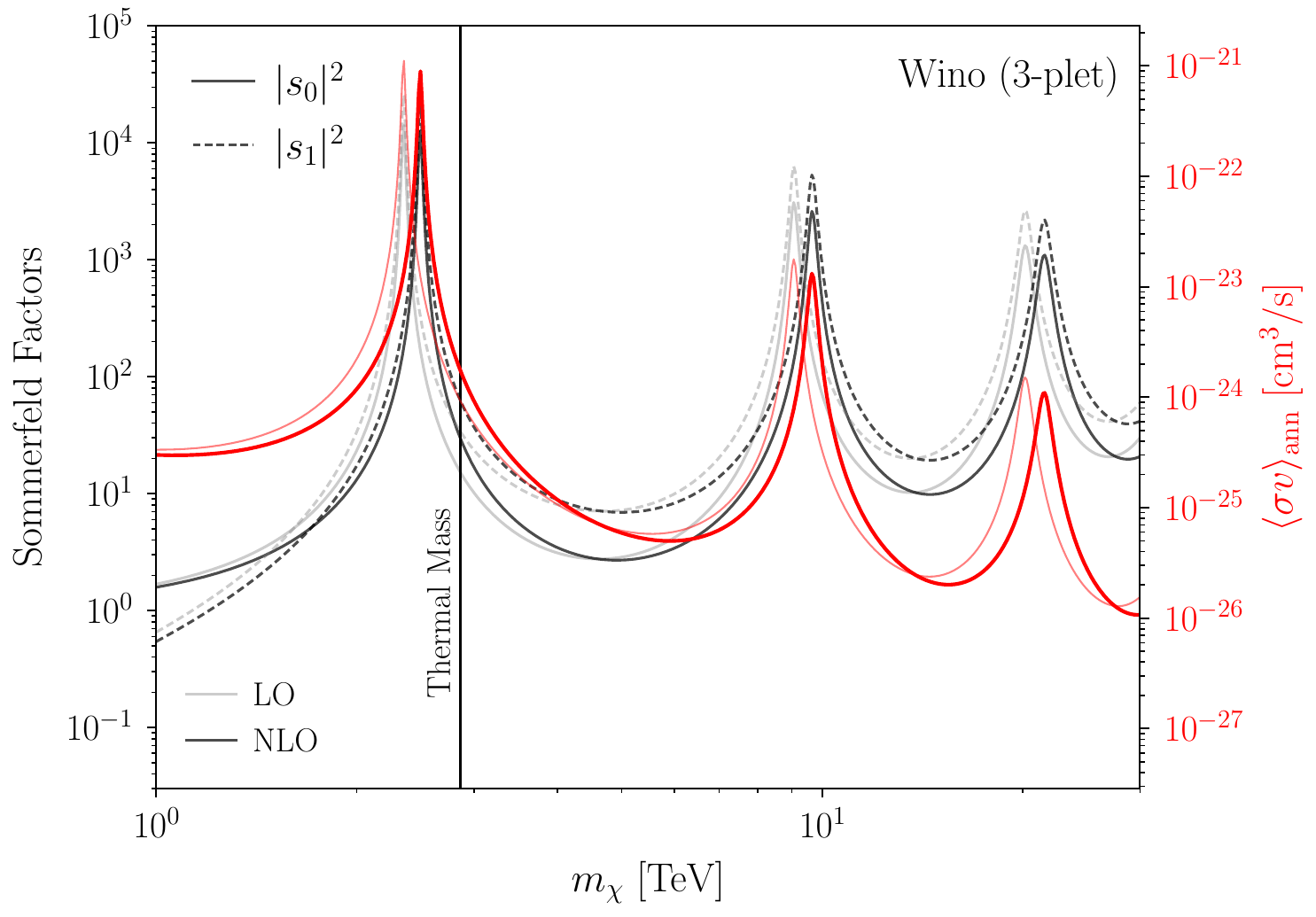}

\includegraphics[width = 0.45\textwidth]{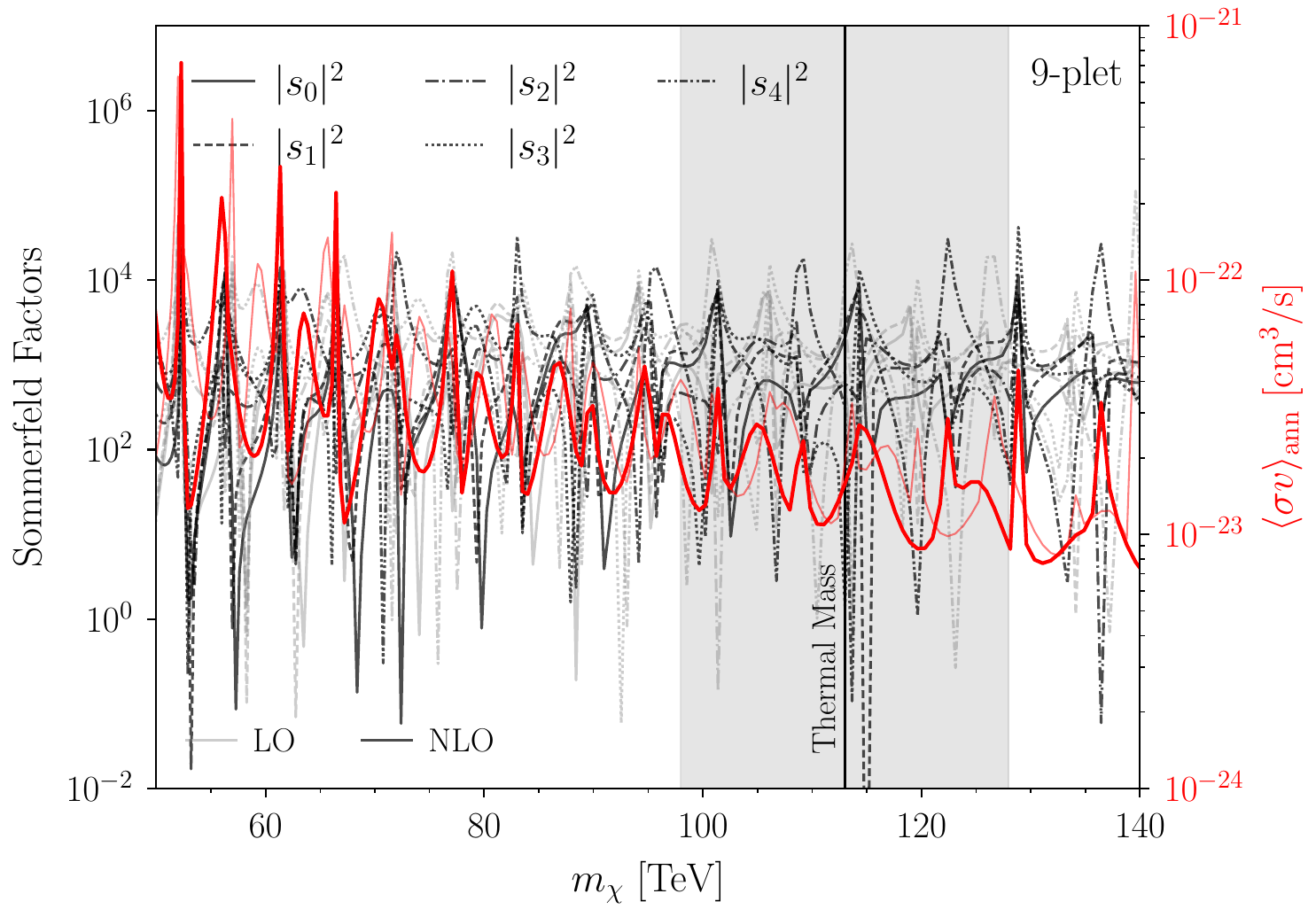}
\caption{Sommerfeld factors (black) and cross sections for direct annihilation (red) for the SU(2) triplet (the wino, top) and 9-plet (bottom). The Sommerfeld resonances occur when states cross the threshold of zero binding energy, and become denser in structure for higher masses and larger representations.  Depending on proximity to these resonances, the expected annihilation for fixed mass can be violently altered by incorporating NLO terms to the EW potential (darker vs lighter lines).} 
\label{fig:Sommerfeld}
\end{figure}

We solve these equations numerically using the Variable Phase Method~\cite{Ershov:2011zz}, which was adapted to DM contexts in~\cite{Asadi:2016ybp,Beneke:2014gja}. The enhancement of the $s$-wave  annihilation cross section from this effect is encapsulated in the Sommerfeld factors $s_i \equiv \psi_i(0)$, where again the wavefunction element $\psi_i$ corresponds to the 2-body state $\chi^{i}{\bar \chi}^{i}$,  
\begin{equation}
    \langle \sigma v \rangle_{\rm ann} = \sum_{GG'} \sum_{ij} s_i \Gamma^{GG'}_{ij} s_j^*,  
\end{equation}
and $\Gamma_{ij}$ is the annihilation matrix element for states  $\chi^i {\bar \chi}^j$ into SM gauge boson final states $G G'$ ($WW, ZZ, Z\gamma, \gamma\gamma$). In this work, we take these quantities at tree level, which dominates the continuum spectrum we seek to study. Figure~\ref{fig:Sommerfeld} illustrates the Sommerfeld factors (norm-squared) and resultant cross section for direct annihilation at halo velocities $v\sim 10^{-3}$ of the Wino (top) and 9-plet (bottom). The resonance structure of these factors become highly intricate and dense  at higher representations, especially compared to the increasing theory uncertainty of the thermal mass, which then imparts significant uncertainty in the expected present-day signal in our Galaxy.    

\subsection{Bound state formation and decay}

At higher representations and larger DM masses, the electroweak potential becomes deep enough to support a large number of bound states. A significant contribution to the depletion of DM then comes from the capture of incoming particles into unstable bound states, and the subsequent decay of these states into visible particles. This family of depletion channels is also a significant contributor in setting the thermal mass for  representations with $n>3$~\cite{Bottaro:2021snn,Mitridate:2017izz}. For this work we consider only the capture processes that involve emission of a single gauge boson --  $\gamma, Z$, or $W^\pm$. We follow heavily the formalism outlined in~\cite{Baumgart:2023pwn, Asadi:2016ybp}, extended to higher-$n$ representations.

The cross section of these capture processes is set by the overlap between the scattering and bound-state wavefunctions. The former of these are obtained via solving~\eqref{eq:sf} for the pure-$\chi^0$ plane-wave boundary condition and appropriate quantum number $L$. In the dipole approximation, the emission of a single gauge boson induces a change $\Delta L = 1, \, \Delta Q=0$ for $\gamma$ or $Z$ emission, and $\Delta L =1, \, \Delta Q = 1$ for $W$ emission. Since the spin $S$ of the two-particle state is unaltered, and the incoming state is always $\chi^0\chi^0$ with $L+S$ even and $Q=0$, the bound states we consider will all have $L+S$ odd and $Q \in \{0, \pm1\}$. For simplicity, we will only account for $s\to p$ and $p \to s$ transitions in computing the capture cross section in this work, which is expected to encompass the dominant contribution, but is in any case strictly conservative. 

The wavefunctions of negative-energy bound states can thus be notated as $\phi^{Ni}_{LQ}$, where $N$ labels the depth of the state in the spectrum, and $i$ span the allowed 2-body states: $\chi^i{\bar \chi}^i$ for $Q=0$, and eg. $\chi^{i+1} {\bar \chi}^{i}$ for $Q = +1$, one entry shorter than the vector of neutral states. They are eigenstates of the equations, 
\begin{equation}
   \frac{1}{m_\chi} (\phi^{Ni}_{LQ})'' - \left( V^{ij}_{Q} + \frac{L (L+1)}{m_\chi r ^2}\right)\phi_{LQ}^{Nj} = E^N_{LQ} \phi_{LQ}^{Ni},
\end{equation}
where the derivative here is respect to $r$ and  $V^{ij}_{Q} $ is described by~\eqref{eq:Vew_q0} for the $Q=0$ case.  In practice, these eigenstates are approximately constructed via (finite) linear combinations of eigenstates for the equivalent equation with unbroken electroweak symmetry, for which the analytic form is known. We estimate the true bound state wavefunction via a truncated spectrum of the first 50 unbroken $SU(2)$ states of corresponding $L, m$ and $Q$, though we find that the smaller electroweak representations converge within the first 10-30. A detailed description of this technique can be found in Ref.~\cite{Asadi:2016ybp}.

As alluded to before, the capture cross section is set by the overlap between the scattering and bound cross sections.  The gauge boson emitted in the capture can originate from either an external fermion leg or the gauge boson exchange within the potential, and both must be accounted for.  For capture via photon emission for instance,  the cross section for capturing from a scattering state with $L$ into the $N^{\rm th}$ neutral bound state of $L'$ (with $|L-L'|=1$) is given by
\begin{align}
    \sigma v|^{\gamma}_{NL'} & = \frac{2 \alpha k_N}{\pi m_\chi^2} \int \Omega_k  \left| \epsilon (\hat k_N) \cdot \int d^3 r \left[ \sum_{i=1}^{(n-1)/2}   i \psi^*_{L,i} \nabla \phi^{N,i}_{L'0} \right. \right. \nonumber\\
    & \left. \left. +\frac{\alpha_W m_W}{2} \hat r \; {\rm e}^{-m_W r} \left( \sqrt{2} C_{W,0}^2  \psi^*_{L,1} \phi^{N,0}_{L'0}\phantom{\sum_{i=1}^{(n-1)/2}}  \right. \right. \right. \nonumber\\
    & \left. \left. \left. + \sum_{i=1}^{(n-1)/2}  C_{W,i}^2 \psi^*_{L,i+1} \phi^{N,i}_{L'0} - C_{W,i}^2\psi^*_{L,i} \phi^{N,i+1}_{L'0} \right)  \right]  \right|^2, 
\end{align}
where $k_N$ is the momentum carried away by the emission, set by the binding energy of the state. The emission via $Z$ boson is essentially identical, except with the prefactor coupling scaled by $c_W^2/s_W^2$ and a modified $k_N$ due to the boson mass. The capture via emission of $W$ can be similarly written and is detailed in Ref.~\cite{Baumgart:2023pwn} for the 5-plet.

\begin{figure}[!htb]
\includegraphics[width = 0.45\textwidth]{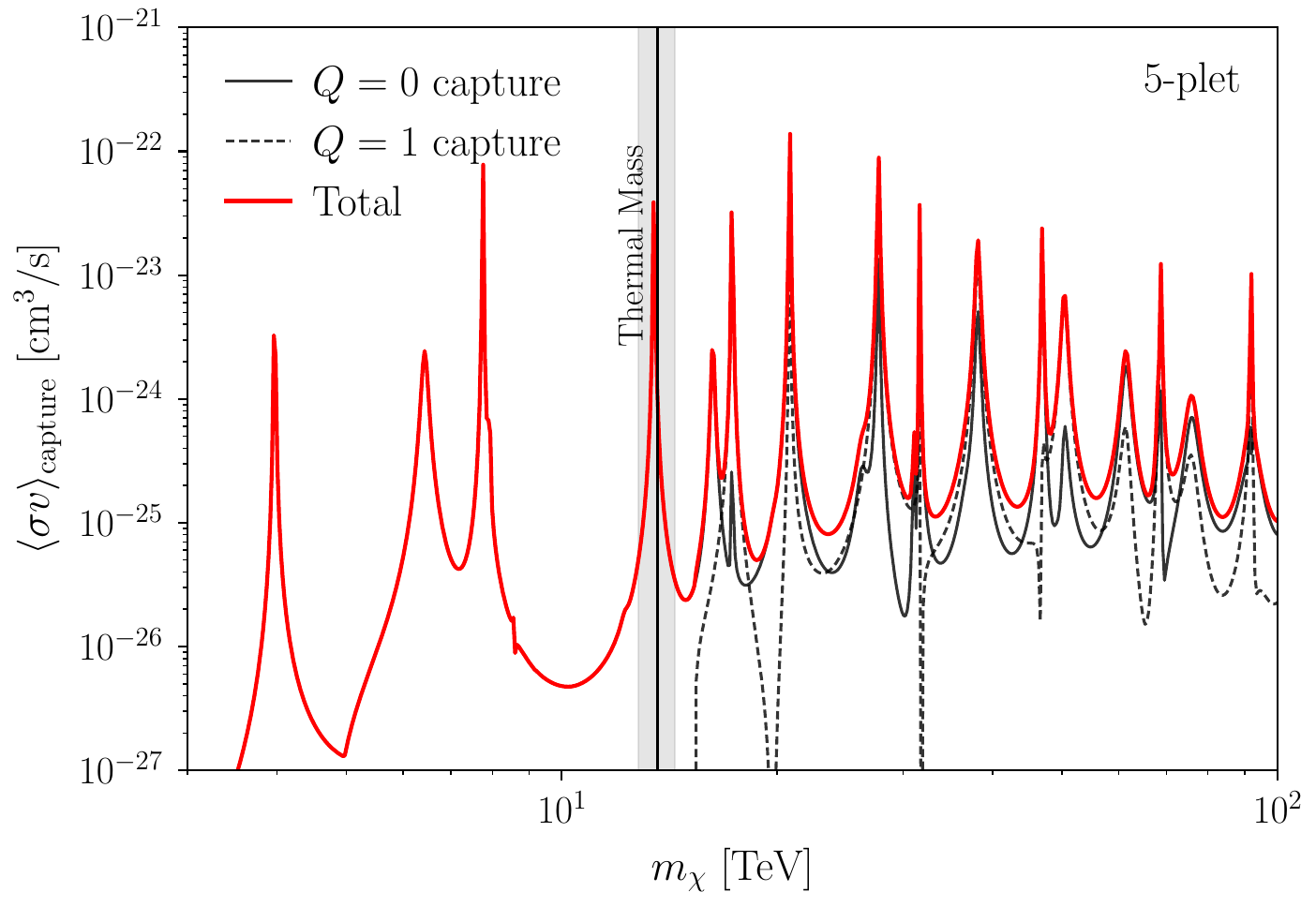}

\includegraphics[width = 0.45\textwidth]{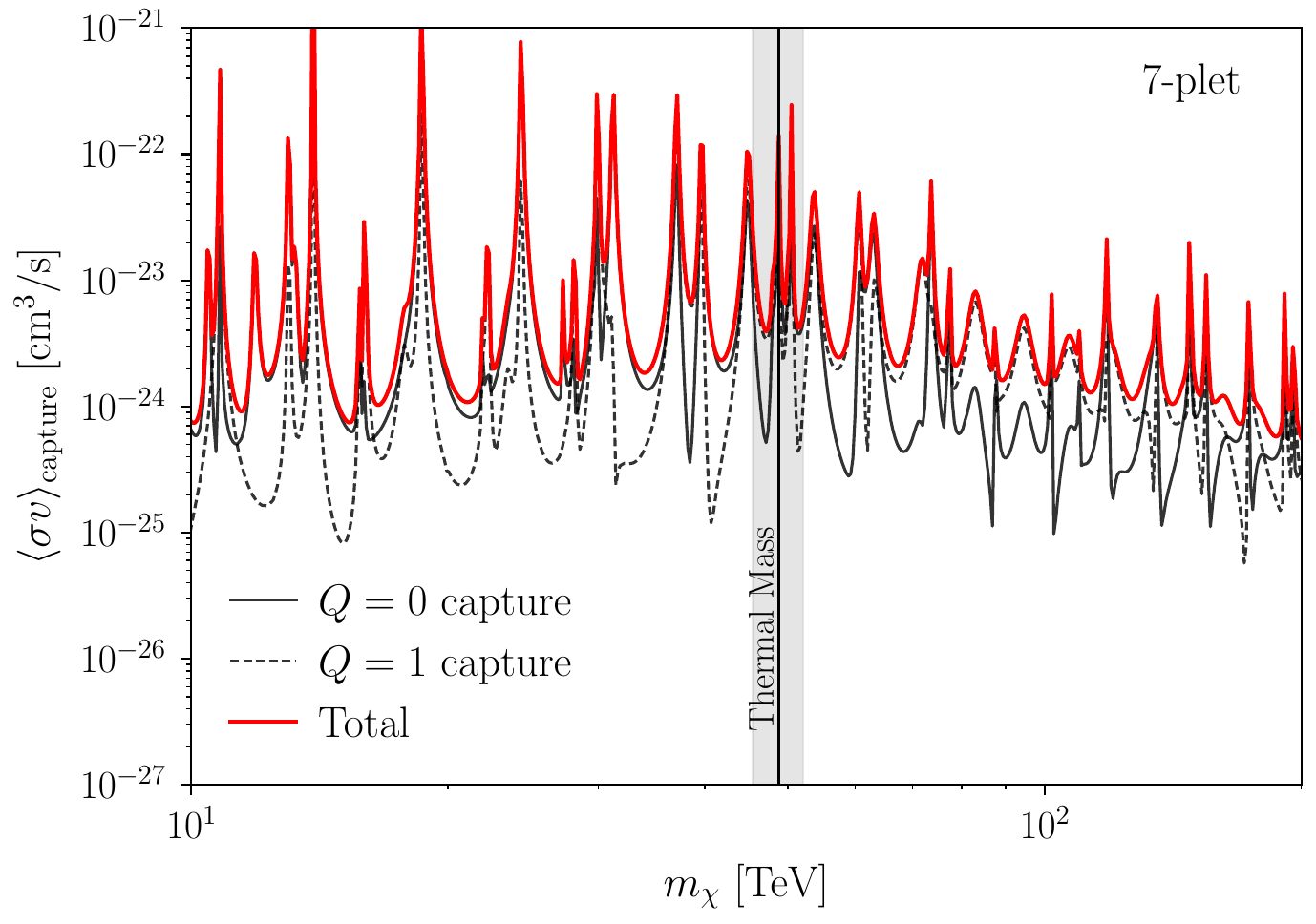}
\caption{Cross section corresponding to bound state capture for the 5-plet (top) and 7-plet (bottom), accounting for single emission of $\gamma/Z$ ($Q=0$) or $W$ ($Q=1$) boson, and including $s\to p$ and $p \to s$ capture processes.}
\label{fig:bsf}
\end{figure}

\begin{figure*}[!htb]
\includegraphics[width = 0.45\textwidth]{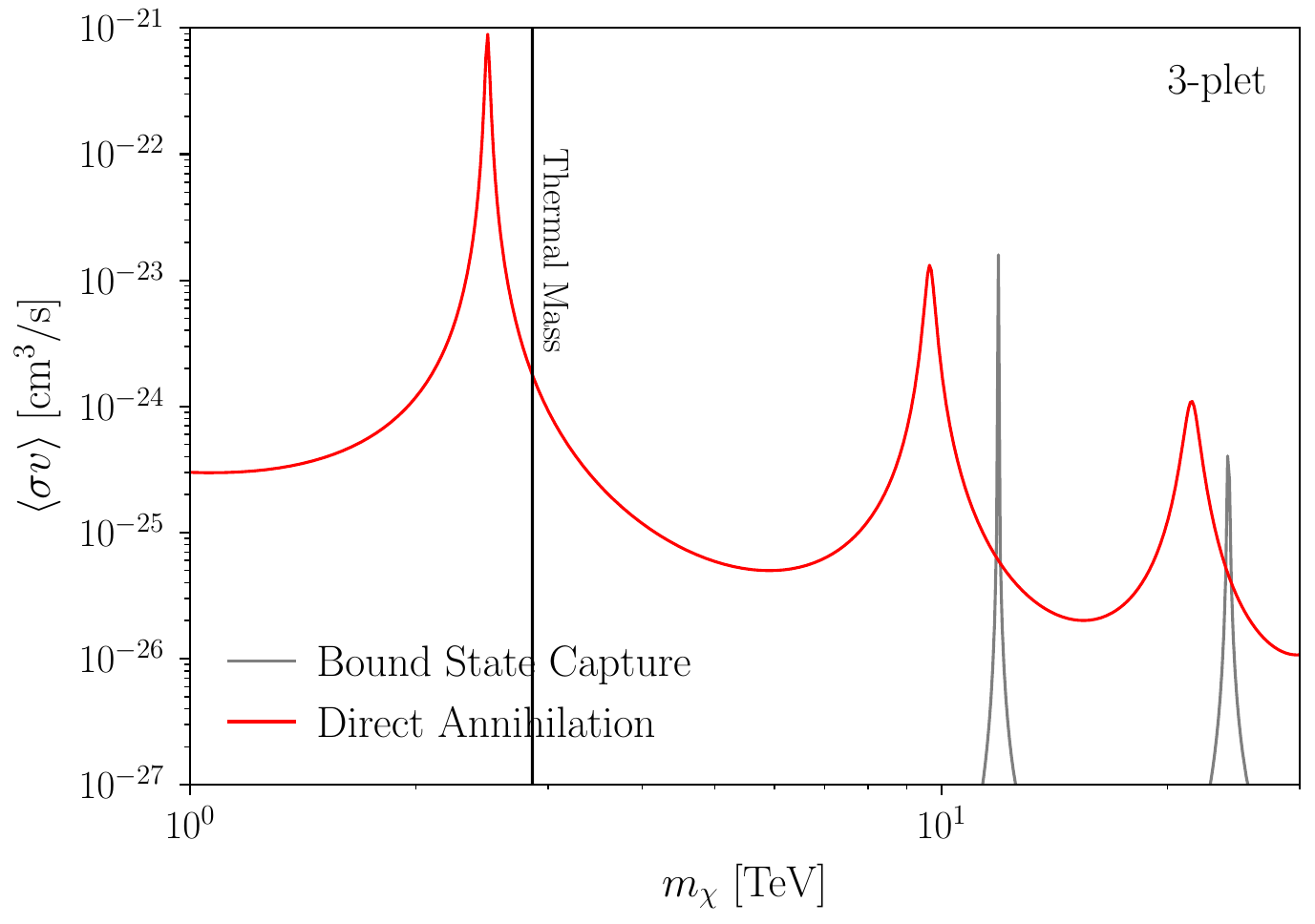}
\includegraphics[width = 0.45\textwidth]{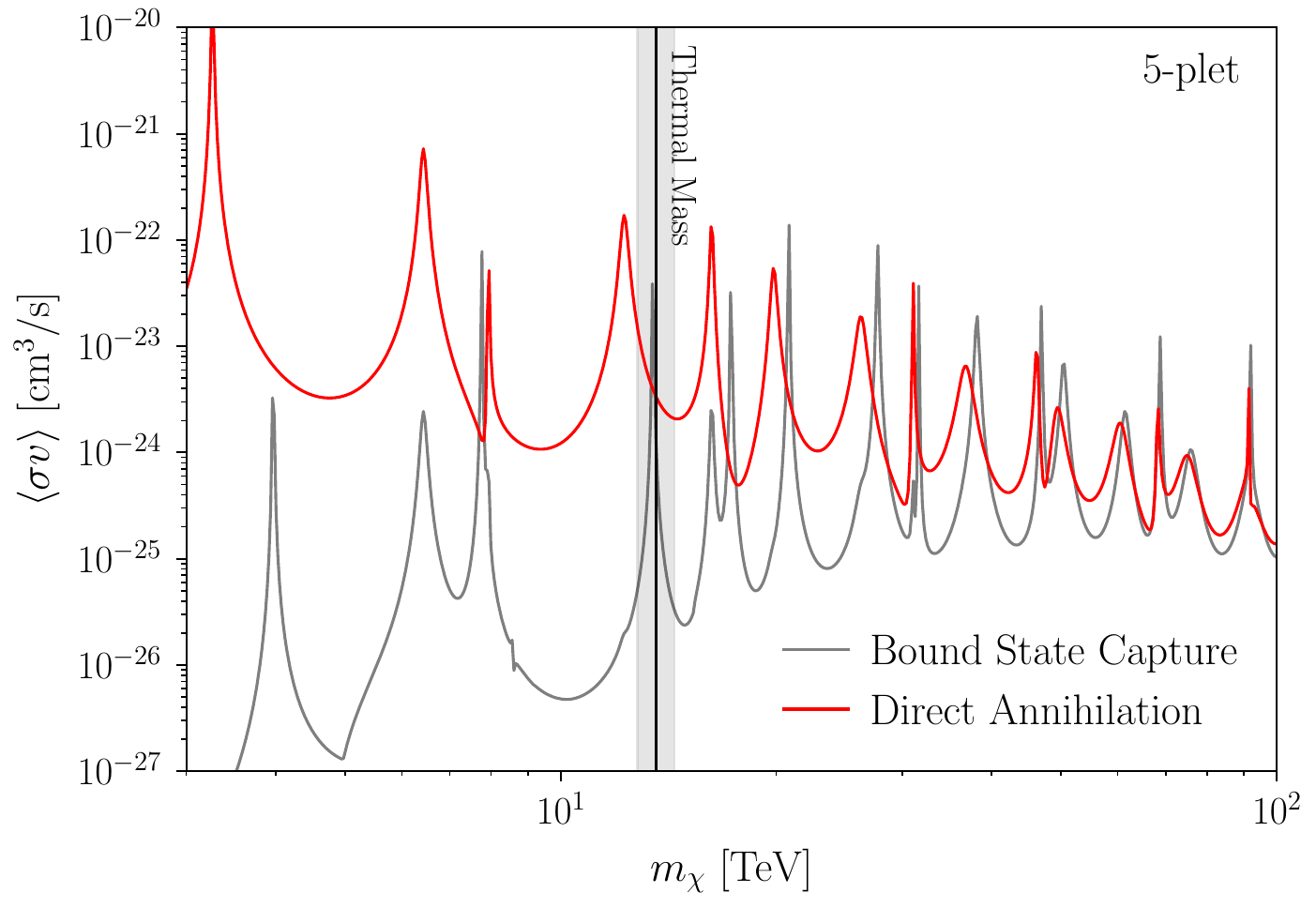}

\includegraphics[width = 0.45\textwidth]{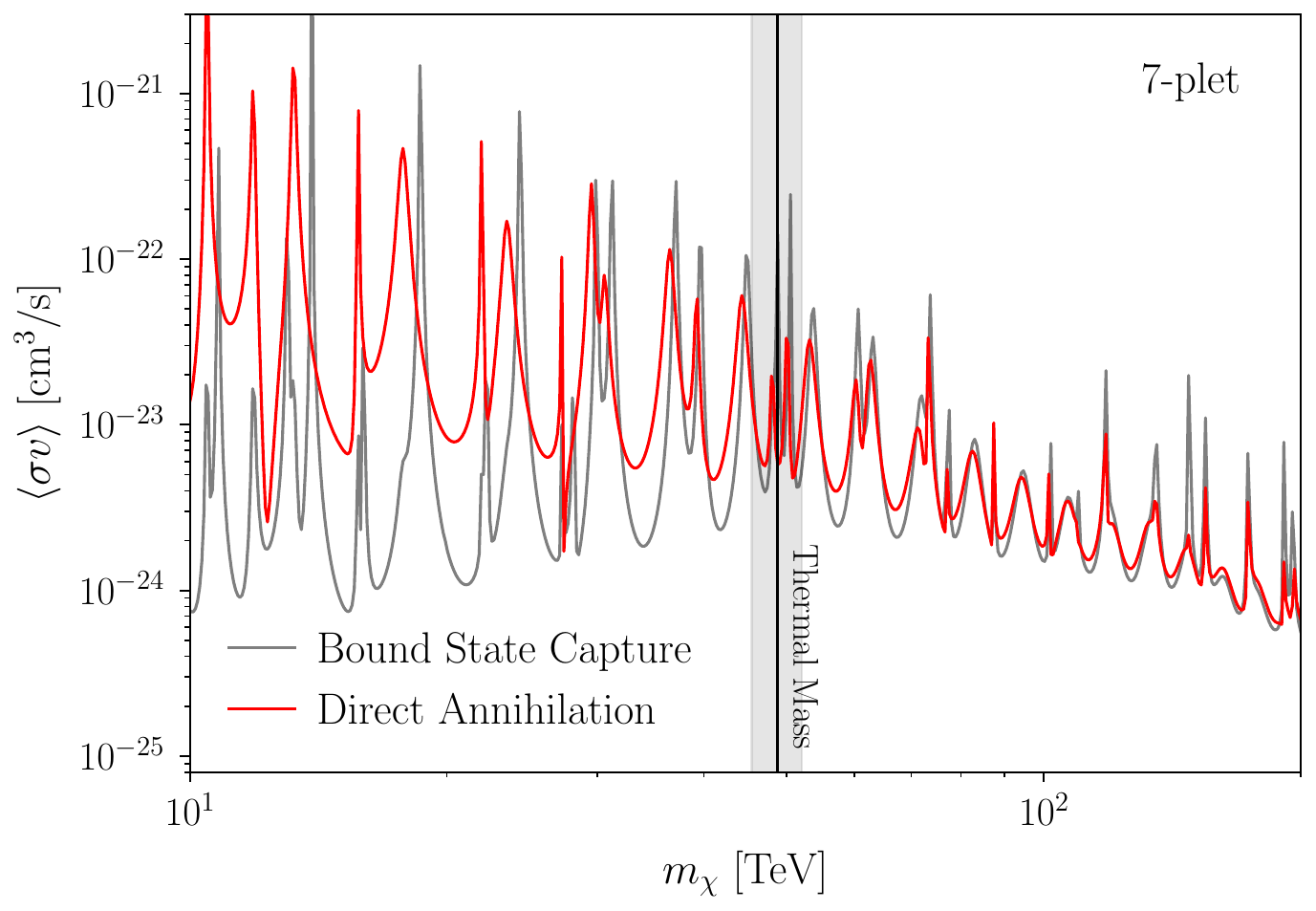}
\includegraphics[width = 0.45\textwidth]{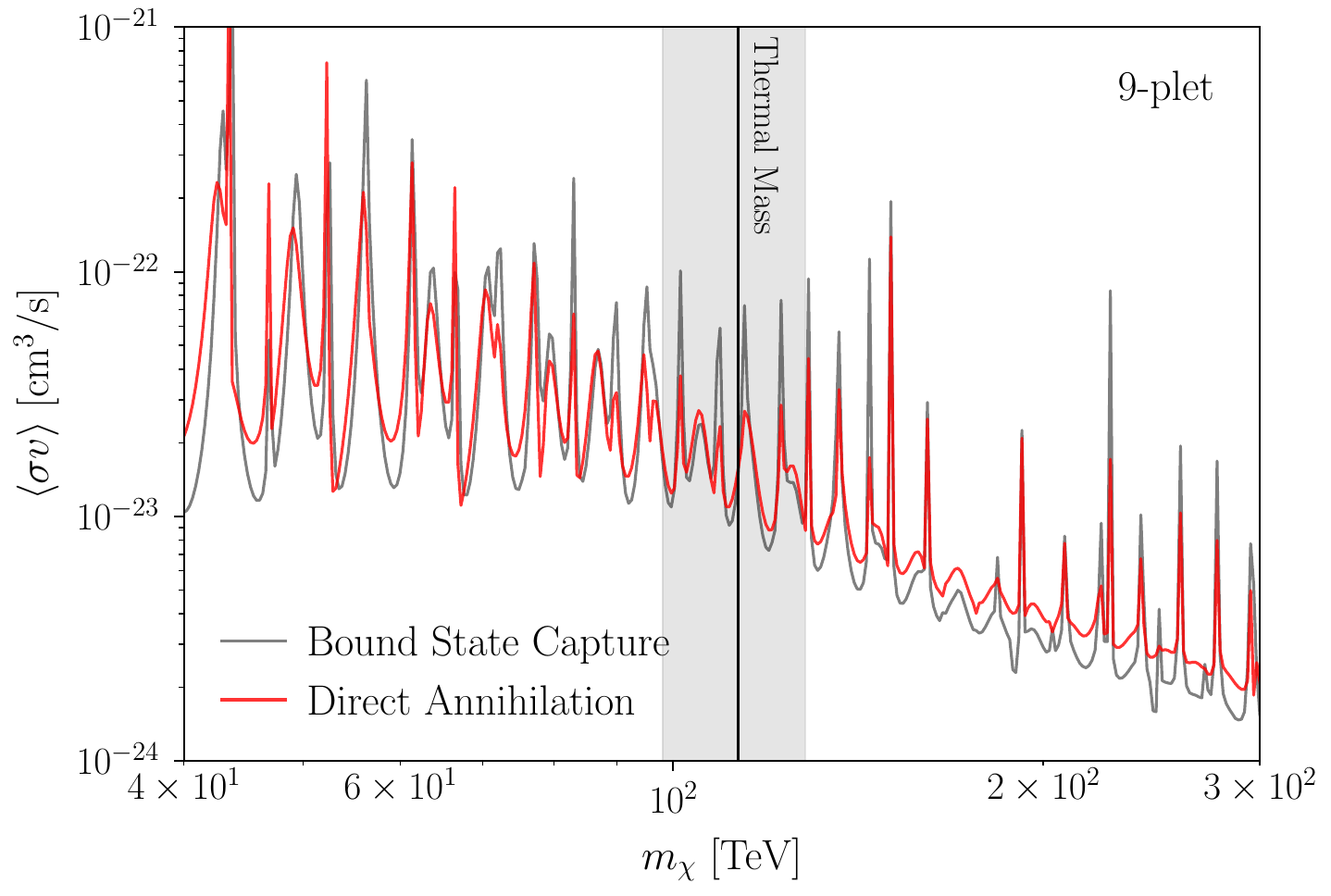}
\caption{Cross section contributions from direct annihilation (red) and bound state capture (black) at NLO for the four real electroweak representations we consider in this work. The bound state contribution to DM depletion is entirely absent for wino masses close to the thermal value, and is only relevant for the quintuplet due to the accidental proximity of the thermal mass to a resonance. For the  7-plet and 9-plet however the capture and direct annihilation are very much  comparably efficient processes in their respective thermal mass ranges.}
\label{fig:xsec_comp}
\end{figure*}

Figure~\ref{fig:bsf} illustrates the bound state capture cross sections for the quintuplet (top) and 7-plet (bottom), broken down into contributions from charged and neutral gauge boson emission, including both $s\to p$ and  $p\to s$ capture. Figure~\ref{fig:xsec_comp} further illustrates the total depletion cross section, both from direct annihilation and bound-state capture, for the four minimal candidates we focus on in this paper.  As expected, the bound state aspect becomes increasingly important for the larger representations.

Finally, after the bound state is formed, it may undergo a decay cascade into increasingly deeper bound states via emitting additional gauge bosons, until it finally into SM particles. While the details of this cascade is important for computing the endpoint of the spectrum, the continuum spectrum is significantly less sensitive, as long as the bound states are indeed unstable. Thus, we compute only the capture into the initial bound state, and model the gamma ray injection of that process as corresponding to decay mediated by an off-shell $Z$ or $W$. 

\subsection{Endpoint of the wino annihilation spectrum}

Since, in this work, we are interested in the gamma rays accessible to the  Fermi telescope, and most of these candidates have thermal masses above 2 TeV, the signal gamma rays we concern ourselves with are almost exclusively generated by the continuum emission from $WW$ and $ZZ \, (+ \frac{1}{2} Z\gamma)$ channels. The exception to this is the wino, whose thermal mass is just outside the Fermi sensitivity window, and for which the search for $\mathcal{O}(0.1- 1) \text{ TeV}$ candidates (which may or may not constitute 100\% of the observed abundance), are an area of mutual interest for collider and indirect probes.  

The Sommerfeld-enhanced annihilation of Winos to line-like gamma rays, $\chi^0\chi^0 \to \gamma\gamma\, (+\frac{1}{2} \gamma Z)$, is very simply related to the $ZZ$-channel continuum contribution, as $\langle \sigma v \rangle_{\gamma} = s_W^2/c_W^2 \langle \sigma v \rangle_{Z} $.  We caution however, that this does not encompass effects that arise as particular consequences of interrogating the hard photon endpoint of the spectrum, where resummation of large logarithmic corrections considerably modify the strength of the line-like signal. Details of this calculation, necessary for the accurate prediction of line signals, are discussed in Refs.~\cite{Bauer:2014ula, Baumgart:2014vma,Ovanesyan:2014fwa,Beneke:2019qaa,Baumgart:2018yed,Baumgart:2023pwn}. Logarithmic electroweak corrections from the hierarchy of scales between the DM and gauge boson masses, {\it i.e.} from the Sudakov factors $\sim \log^2 (m_\chi/m_W)$, are entirely agnostic to DM representation and are encapsulated, for our prediction of the continuum emission, in the injection spectra piece $dN_{WW (ZZ)}/dE$ of~\eqref{eq:ann_formula}; those details are discussed in Refs.~\cite{Cirelli:2010xx,Bauer:2020jay}. In our work when computing the line-like signal for the wino at the spectral endpoint we use the {\tt DM}$\gamma${\tt Spec} package presented in~\cite{Beneke:2022eci}, for which all of the previously-mentioned effects are incorporated. 

\section{Fermi Continuum data analysis}
\label{sec:data}

We search for the continuum spectra from DM annihilation from the wino and higher-dimensional-representation minimal DM models using Fermi gamma-ray data. We use the same data set as used in Ref.~\cite{Dessert:2022evk}, consisting of 722 weeks of \texttt{SOURCE} data selected for the top 3 of 4 quartiles of data as ranked by the point spread function (PSF), in order to minimize cosmic-ray-induced backgrounds and contamination from point sources (PSs). We apply the standard quality cuts to the data selection (see~\cite{Dessert:2022evk} for details), and we also implement the 4FGL~\cite{Fermi-LAT:2019yla} PS mask constructed in~\cite{Dessert:2022evk}.  We mask the Galactic plane to only consider latitudes $|b| \geq 1^\circ$, and we consider 19 Galactocentric radii of radial width $1^\circ$, with the first annulus $1^\circ - 2^\circ$ and the 19$^{\rm th}$ from $19^\circ - 20^\circ$. (See Ref~\cite{Dessert:2022evk} for an image of the data.)
  
\begin{figure}[!htb]
\includegraphics[width = 0.49\textwidth]{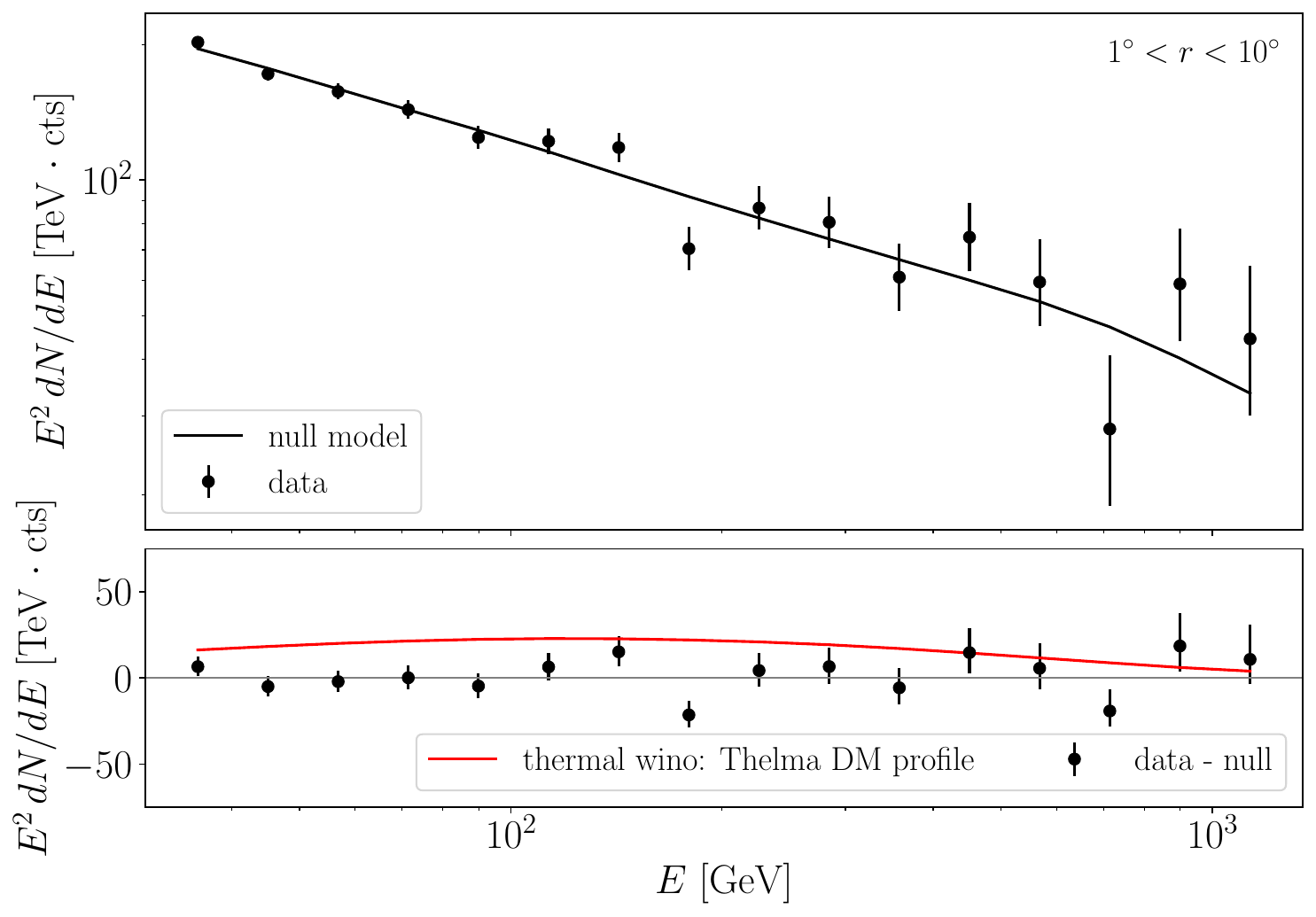}
\caption{ (Top) An illustration of the {\it Fermi} spectral data in the inner Galaxy used in this work to search for minimal DM. In our fiducial analysis we analyze the data in 19 annuli independently, of width $1^\circ$ starting with the annulus from $1^\circ$ to $2^\circ$, but in an alternate analysis we analyze the spectral data summed over the inner $10^\circ$ (illustrated here), subject to a Galactic plane and PS mask.  The null hypothesis model is the \texttt{p8r3} Fermi diffuse spectral model with a nuisance parameter for its overall normalization---the best-fit null model is illustrated for the alternate analysis. (Bottom) Residuals under the null hypothesis for the alternate null-hypothesis analysis. On top of the residuals we show the expected contributions from the thermal wino assuming the Thelma DM density profile, which is the most conservative of the FIRE-2 Milky Way analogue density profiles.  }
\label{fig:Fermi_data}
\end{figure} 
The data are binned in energy into 40 logarithmically-spaced energy bins between 200 MeV and 2 TeV; however, in our analysis we include data between and including energy bins 23 and 37 ($\sim$31.7 GeV and $\sim$1262 GeV).  The spectral data are illustrated in Fig.~\ref{fig:Fermi_data} summed over the first 9 annuli, corresponding to angles less than 10$^\circ$ from the Galactic Center.  

For our fiducial analysis we model the spectral data in each annulus independently using a combined signal and background model ${\mathcal M}$ with likelihood 
\es{eq:L}{
p({\bm d} | {\mathcal M}, {\bm \theta}) = \prod_{i,j} {\mu_{i,j}({\bm \theta})^{N_{i,j}} e^{-\mu_{i,j}({\bm \theta})} \over N_{i,j} !} \,,
}
where ${\bm d} = \{ N_{i,j} \}$ is the set of counts across annuli $i$ and energy bins $j$.  The model parameter vector has components ${\bm \theta} = \{ \mu, {\bm \theta}_{\rm nuis.} \}$, with $\mu$ the signal model parameter, discussed more later in this section, parameterizing the annihilation cross-section and ${\bm \theta}_{\rm nuis.}$ the set of nuisance parameters that govern the null-hypothesis model.  The null-hypothesis model, which is nested in the signal plus background model with $\mu = 0$, has a single nuisance parameter per annulus, which we refer to as ${\theta}_{\rm nuis}^i$. The signal plus background model prediction in annulus $i$ across all energy bins is then
\es{}{
\mu_{i,j}({\bm \theta}) = \mu S_{i,j}^{\rm sig} + {\theta}_{\rm nuis.}^i B_{i,j}^{\rm bkg} \,,
}
where $S_{i,j}^{\rm sig}$ and $B_{i,j}^{\rm bkg}$ are the signal and background model spectral templates in the respective annuli and energy bins. 
 We describe these templates in more detail below.  For now, we note that given~\eqref{eq:L} we profile over the nuisance parameters to construct the profile likelihood $\lambda(\mu)$, which is defined by
\es{eq:pL}{
\lambda( \mu ) \equiv { p({\bm d}|{\mathcal M}, \{ \mu , \hat {\hat {\bm \theta}}_{\rm nuis.}\}) \over p({\bm d}|{\mathcal M}, {\hat {\bm \theta}}) } \,,
}
where ${\hat {\bm \theta}}$ is the model parameter vector that maximizes the likelihood, and $\hat {\hat {\bm \theta}}_{\rm nuis.}$ is the nuisance parameter vector that maximizes the likelihood at fixed $\mu$. Given the profile likelihood $\lambda(\mu)$, we perform standard frequentist inference---assuming statistics in the asymptotic limit described by Wilks' theorem---to search for evidence and constrain the signal model (see~\cite{Safdi:2022xkm} for a review). In particular, throughout this work we quote 95\% one-sided upper limits on $\mu$ and the discovery test statistic (TS) in favor of the DM model. The 95\% one-sided upper limit on $\mu$ is given by the $\mu > \hat \mu$, with $\hat \mu$ the best-fit value, where $-2 \log \lambda(\mu) \approx  2.71$~\cite{Safdi:2022xkm}.  The discovery TS for the two-sided test, where we allow for $\mu < 0$ in addition to $\mu > 0$, is 
\es{}{
{\rm TS} = 2 \log {\lambda(\hat \mu) \over \lambda (0)} \,.
}
A discovery TS less than 1 (4) implies less than 1$\sigma$ (2$\sigma$) evidence in favor of the two-sided DM model. Of course, in reality we care about one-sided tests, since $\mu < 0$ are unphysical, but we quote results for the two-sided TSs in this work to also assess for mismodeling, which can at times be diagnosed through large TSs with negative $\hat \mu$. 

We construct the background model $B_{i,j}^{\rm bkg}$ using the reprocessed Fermi \texttt{gll\_iem\_v07} (\texttt{p8r3}) Galactic emission model.  This model accounts for standard gamma-ray production in the Galaxy, mostly arising from boosted pion decay from accelerated cosmic ray protons hitting gas within the Galaxy; subdominant contributions include, among other sources, inverse Compton emission from cosmic ray electrons and emission from the Fermi bubbles~\cite{Su:2010qj}.   Our best-fit null model, for the alternate analysis described below, is shown in Fig.~\ref{fig:Fermi_data}, with no obvious signs of mismodeling.   

Our signal model is constructed by forward modeling the minimal DM annihilation spectrum, described in the previous section, through the Fermi instrument response using the exposure map obtained in the data-reduction process. Note that we neglect the $\sim$10\% energy resolution of Fermi in our analysis given our large energy bins and since the broad-spectrum continuum signal does not have structure on small energy scales.  In Fig.~\ref{fig:Fermi_data} we illustrate the wino signal model contribution for an example DM profile, which is discussed in the next section.

Note that the signal model at fixed DM mass and electroweak representation does not have any additional model parameters, since it is completely deterministic. However, we still assign to the signal model a model parameter $\mu$, which is an overall rescaling of the predicted flux with $\mu = 1$ the minimal DM model and $\mu = 0$ the null model. That is, the parameter $\mu$ provides an interpolation between the null hypothesis and the signal hypothesis.  If the 95\% one-sided upper limit on $\mu$ excludes $\mu = 1$, then we say that the minimal DM model is disfavored. 

As an alternate analysis framework, to emphasize the robustness of our search, we simplify the data by summing over the full ROI and modify the likelihood in~\eqref{eq:L} to 
\es{eq:L_alt}{
P({\bm d}_{\rm alt} | {\mathcal M}_{\rm alt}, {\bm \theta}) = \prod_{j} {\mu_{j}({\bm \theta})^{N_{j}} e^{-\mu_{j}({\bm \theta})} \over N_{j} !} \,,
}
with ${\bm d}_{\rm alt} = \{ N_j \}$ being the counts across energy bins $j$ summed over the inner $10^\circ$ of the Galaxy (first 9 annuli), subject to the Galactic plane and PS masks described above.  We chose to restrict to the inner $10^\circ$ of the Galaxy to avoid the DM annihilation signal being overly diluted. We fit the alternate model ${\mathcal M}_{\rm alt}$ to these data with two model parameters, a nuisance parameter $\theta_{\rm nuis.}$ that normalizes the spectral template for the diffuse emission in this ROI and the signal parameter of interest $\mu$. We illustrate the summed data used in the alternate search in Fig.~\ref{fig:Fermi_data} along with the residuals under the null hypothesis. 

\subsection{DM Density Profile in the Inner Galaxy}
\label{sec:DM}

The constraint we obtain on the signal model parameter $\mu$ is highly dependent on the assumed DM density profile, with more aggressive choices of DM density profile leading to stronger constraints.  There are strong reasons to believe that commonly-assumed DM density profiles, like the Navarro–Frenk–White (NFW) profile~\cite{Navarro:1995iw,Navarro:1996gj} and the Einasto profile~\cite{Graham:2005xx,Retana-Montenegro:2012dbd,Navarro:2008kc}, which are motivated by cold, DM-only $N$-body simulations, do not properly model the DM density in the inner Milky Way because the gravitational potential in this region is dominated by baryons. Baryons can increase the inner DM density through adiabatic contraction but also core the density profile through feedback mechanisms.  To address this issue some previous works have supplemented DM-only profiles like the NFW profile with artificial DM cores up to multiple kpc in width ({\it e.g.},~\cite{Fan:2013faa,Cohen:2013ama,Baumgart:2023pwn}).  On the other hand, a perhaps more principled approach is to use DM profiles from Milky Way analogue galaxies in cosmological hydrodynamic simulations~\cite{Dessert:2022evk,Rodd:2024qsi,Hussein:2025xwm,Abe:2025lci}.  

In this work we take both approaches described above. On the one hand, as we argue below, the lower end of $J$-factors found in the FIRE-2 Milky Way analogue galaxies~\cite{Hopkins:2017ycn,2022MNRAS.513...55M} represents a conservative lower bound on the $J$-factor profile for the Milky Way.  We thus use this lower bound on the $J$-factor, which arises from a particular Milky Way analogue galaxy named `Thelma' in the FIRE-2 simulation suite, when conservatively addressing whether the Fermi continuum gamma-ray data rules out or allows for minimal DM. However, in the context of the wino DM model---which is the most motivated of the real minimal DM scenarios---we additionally ask the question of how extreme the DM density profile would need to be in order for the thermal wino to avoid the constraints of our analysis.  In this context we consider a cored Einasto DM profile, with functional form
\es{eq:cored_einasto}{
\rho_{\rm Ein}^{\rm core}(r) = \rho_s \left\{ 
\begin{array}{cc}
 \exp \left[ - {2 \over \alpha_s} \left( \left( {r_c \over r_s} \right)^{\alpha_s} - 1 \right) \right] \,, & r < r_c \\
 \exp \left[ - {2 \over \alpha_s} \left( \left( {r \over r_s} \right)^{\alpha_s} - 1 \right) \right] \,, & r > r_c \,,
\end{array} \right.
}
where $r_c$ is the core radius and $r$ is the distance from the Galactic Center.
We fix $r_s = 20$ kpc and $\alpha_s = 0.17$ as in~\cite{Safdi:2022xkm,Rodd:2024qsi}.  

All Milky Way DM density profiles, such as the cored Einasto profile in~\eqref{eq:cored_einasto} or the FIRE-2 Milky Way analogue galaxy profiles discussed more below, have an overall normalization parameter that sets both the local DM density and the amount of DM in the Galaxy, both of which are independently measured. The recent review Ref.~\cite{deSalas:2020hbh} concludes that studies of the local stellar velocity dispersion perpendicular to the disk prefer local DM densities in the range $0.4$-$0.6$ GeV/cm$^3$. Here, we follow~\cite{2022MNRAS.513...55M} and set the local DM density to $\rho_\odot = 0.38$ GeV/cm$^3$, which is the lowest allowed value found in~\cite{Guo:2020rcv}, which used LAMOST and Gaia data for stars in the local neighborhood in conjunction with vertical Jeans modeling to determine the local DM density.  However, it should be kept in mind that more global efforts to determine the DM density at the solar radius, which we fix to $r_\odot = 8.23$ kpc~\cite{2022arXiv220412551L}, allow for slightly lower local DM densities, roughly in the range $\rho_\odot \sim 0.3$-$0.5$ GeV/cm$^3$~\cite{deSalas:2020hbh}.  The local and global DM density measurements are each subject to their own systematic uncertainties, such as assumptions of symmetry, equilibrium, and the necessity to model the baryonic components of the Galaxy. However, effectively no study, local or global, has found $\rho_\odot$ below $0.3$ GeV/cm$^3$, and certainly none have found DM densities as low as $0.2$ GeV/cm$^3$, at which point the DM density would likely be much too low to explain measurements of the total mass of the Milky Way~\cite{McMillan:2016jtx}.  Thus, while we adopt $\rho_\odot = 0.38$ GeV/cm$^3$ as a fiducial choice in this work, as an extreme lower bound on the local DM density when considering the wino model we consider the possibility that the local DM density is $\rho_\odot = 0.2$ GeV/cm$^3$.  Note that we determine $\rho_s$ in~\eqref{eq:cored_einasto} through our choice of $\rho_\odot$, and also in the context of the FIRE-2 Milky Way analogue $J$-factors we determine their overall normalizations by requiring that they give the specified local DM density. 

To go beyond phenomenological DM density profile models such as the Einasto or NFW profiles at distances within a few kpc of the Galactic Center, we are forced at the moment to resort to theoretical arguments and simulations, as stellar-kinematic-data-based estimates are unable to resolve the DM density profile in the inner Galaxy at present.  This is not because of the lack of precision of current stellar kinematic data but rather because of the assumptions of equilibrium and uncertainties in understanding the mass of ordinary matter in the inner Galaxy; indeed, within a conservative accounting of uncertainties no amount of DM is required at all to explain the rotation curve within roughly 6 kpc of the Galactic Center, even though DM is definitively required to explain the rotation curve at larger radii (see, {\it e.g.},~\cite{Iocco:2015xga,Benito:2020lgu}). For example, within 3 kpc of the Galactic Center the DM mass fraction relative to the total mass in this region, assuming a cuspy DM profile such as the Einasto profile, is only expected to be of order ${\mathcal O}(10\%)$, which is certainly lower than the precision of the total baryonic mass measurement in this region~\cite{Eilers:2019gqs}. 

Of course---even if allowed by current rotation-curve data---there is no physical reason to expect the DM density of the Milky Way to drop to zero in the inner few kpc of the Galaxy. Cold, DM-only cosmological simulations support cuspy DM profiles in the inner regions of galaxies like the Milky Way, 
and the amount of coring baryonic feedback can provide in the inner region is finite.  These DM cores develop when non-adiabatic processes such as supernovae inject energy into the baryons, causing the baryonic mass distribution to expand.  The DM, which previously was in gravitational equilibrium with the baryons, now has too high of a velocity dispersion and partially evaporates, thus coring the DM density. For this to be efficient the energy injected by feedback mechanisms such as supernovae needs to be sufficiently large to flatten the DM profile, which becomes increasingly hard at large radii because of the lower star formation rates away from the Galactic Center and the fact that the DM distribution at large radii becomes less responsive to impulsive changes to the baryonic distributions. Core sizes from feedback may extend to perhaps at most $\sim$2 kpc~\cite{Butsky:2015pya,2020MNRAS.497.2393L}, which is roughly the extent of the Galactic bulge.  Moreover, as emphasized more below, baryonic effects are expected to make the DM density profile less cuspy but not completely flat, and do not necessarily result in less abundant DM in the inner Galaxy compared to DM-only scenarios.

\begin{figure}[!t]
\centering
\includegraphics[width=\linewidth]{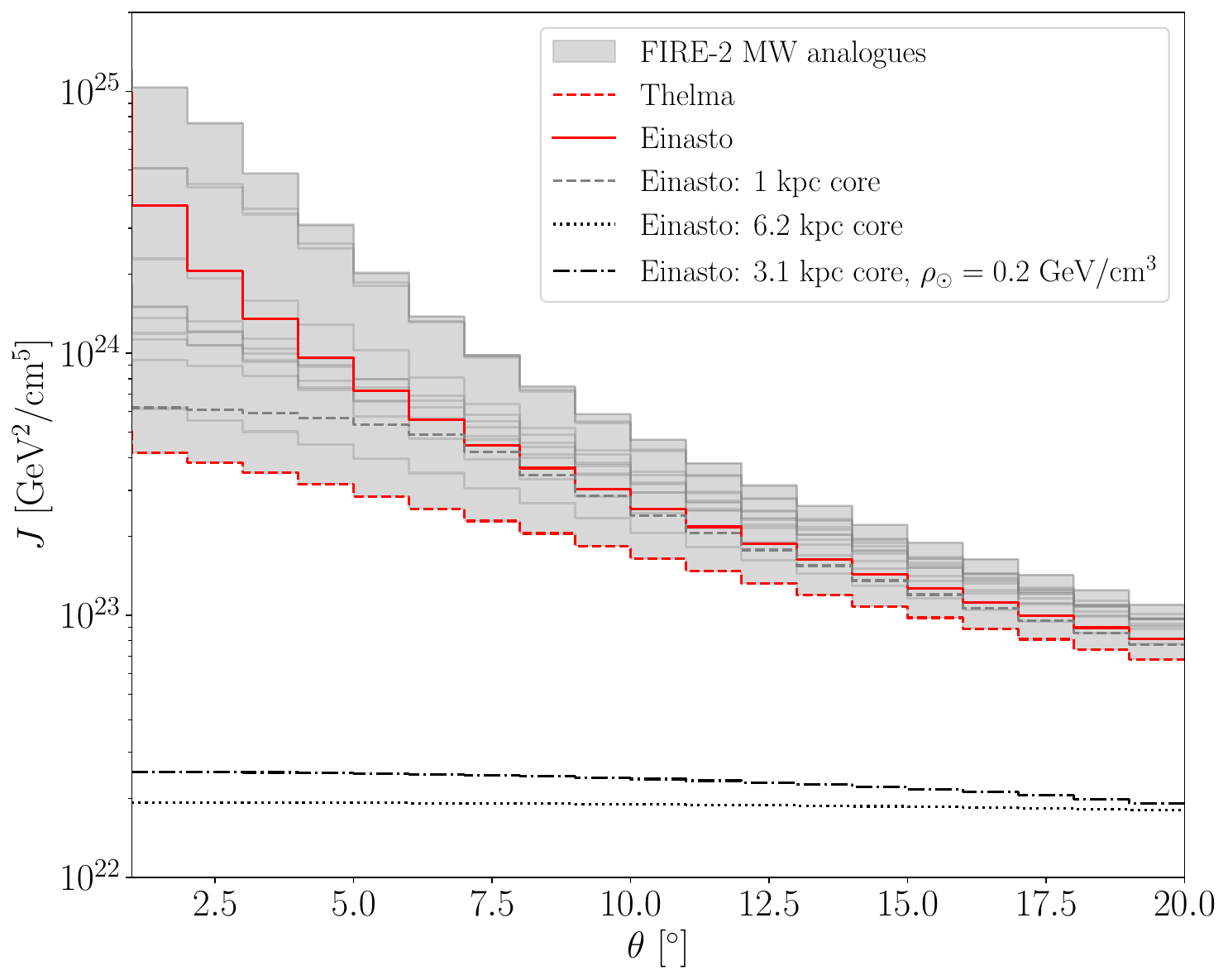}
\vspace{-0.7cm}
\caption{Examples of $J$-factor profiles considered in this work. In light gray we show the 12 Milky Way analogue galaxies from the FIRE-2 simulation suite, which is the simulation suite with the burstiest and strongest feedback mechanisms over all modern simulation suites, leading to conservative $J$-factor estimates relative to those from other simulations~\cite{Hussein:2025xwm}.  As benchmark models in this work we take the Thelma FIRE-2 Milky Way analogue galaxy, which is that with the lowest $J$-factor profile of the 12 galaxies, in addition to the canonical Einasto profile, which is also illustrated. Note that we normalize these profiles to have a common local DM density of $\rho_\odot = 0.38$ GeV/cm$^3$. We additionally show three cored Einasto profiles, whereby the DM density is constant at distances $r < r_c$ from the Galactic Center. The 1 kpc core size is reasonable, but the 3.1 kpc and 6.2 kpc cores are well outside of the ranges expected from baryonic feedback mechanisms. Note that, unlike for the other profiles, we normalize the 3.1 kpc core model to have a local DM density of $\rho_\odot = 0.2$ GeV/cm$^3$.  We are able to rule out all of these DM profiles assuming the thermal wino model in our Fermi data analysis, which we use as strong motivation for claiming that the thermal wino is excluded as a particle of nature explaining 100\% of the DM. }
\vspace{-0.5cm}
\label{fig:J}
\end{figure}

Ref.~\cite{Hussein:2025xwm} recently attempted to bound the $J$-factor profile in the inner Milky Way using the FIRE-2~\cite{Hopkins:2017ycn,2022MNRAS.513...55M,Wetzel:2022man}, Auriga~\cite{Grand:2024xnm}, TNG50~\cite{2024MNRAS.535.1721P}, and VINTERGATAN-GM~\cite{2021MNRAS.503.5826A,Rey:2022mwh} cosmological hydrodynamic simulation suites.  Adiabatic contraction, whereby the baryons in the inner galaxy slowly contract due to dissipation, contracting also the DM distribution, is observed in all of the simulation suites. On the other hand, FIRE-2 has an especially bursty and strong baryonic feedback prescription relative to the other simulation suites, which results in more cored profiles.  Indeed, the other simulation suites only find cuspy DM profiles in their Milky Way analogue galaxies, which are as, or even more, DM-abundant than the Einasto profile across the full ROI of our analysis.  

We thus use the FIRE-2 Milky Way analogue galaxies in order to assess the uncertainty on the DM density profile from baryonic feedback. There are 12 such galaxies (see also discussions in~\cite{Dessert:2022evk,Rodd:2024qsi,Abe:2025lci}), and their $J$-factors are illustrated in Fig.~\ref{fig:J} by the faint gray curves over our ROI and averaged over the individual annuli, starting at $1^\circ$ from the Galactic Center.  In the innermost annulus the spread in $J$ factors among the 12 simulations is over an order of magnitude, though the difference is much smaller at large radii. The resolution of these numerical simulations is estimated to be $\sim$$2^\circ$ at the center of the Galaxy, and is not expected to be a factor of concern for our analysis, which encompasses the inner $20^\circ$. To be conservative we consider Thelma as a benchmark profile throughout this work, the FIRE-2 Milky Way analogue galaxy which gives the lowest $J$-factor~\cite{Hopkins:2017ycn,2022MNRAS.513...55M,Wetzel:2022man} (see Fig.~\ref{fig:J}). We note that this Milky Way analogue galaxy actually has the lowest predicted $J$-factor profile over all Milky Way analogue galaxies in modern hydrodynamic cosmological simulations~\cite{Hussein:2025xwm}.

In Fig.~\ref{fig:J} we also show the Einasto profile with no core, which we use as a secondary benchmark model throughout this work, and three examples of cored Einasto profiles that we consider in the context of the wino. The Einasto cored profile with a 1 kpc core is seen to be within the scope of $J$-factors extracted from the FIRE-2 simulations and thus a reasonable approximation for the level of core that may be expected from baryonic feedback.  As we discuss further in the next section, assuming thermal wino DM our continuum Fermi data analysis is able to rule out Einasto profiles with cores up to $\sim$6.7 kpc, including the 6.2 kpc core profile shown in Fig.~\ref{fig:J}. Even if the local DM density is only $0.2$ GeV/cm$^3$, we are still able to rule out the 3.1 kpc core profile shown in Fig.~\ref{fig:J}.

\section{Wino results}
\label{sec:wino}

We search for continuum emission from wino DM annihilation for wino masses from $m_\chi = 2$ TeV to $m_\chi = 5$ TeV, even though the thermal wino mass is $m_\chi = 2.86 \pm 0.01$ TeV (see Tab.~\ref{tab:MDM}). Below 2 TeV, as we discuss below, the search for a gamma-ray line at the wino endpoint is more constraining than the continuum search, since Fermi is sensitive to gamma-rays up to 2 TeV. We also discuss, later in this section, wino constraints from a variety of other probes, including Fermi searches for continuum gamma-rays from nearby dwarf galaxies and galaxy clusters and searches for Galactic Center gamma-ray lines with ground-based pointing instruments such as HESS and MAGIC.

\begin{figure*}[!htb]
\centering
\includegraphics[width=0.49\linewidth]{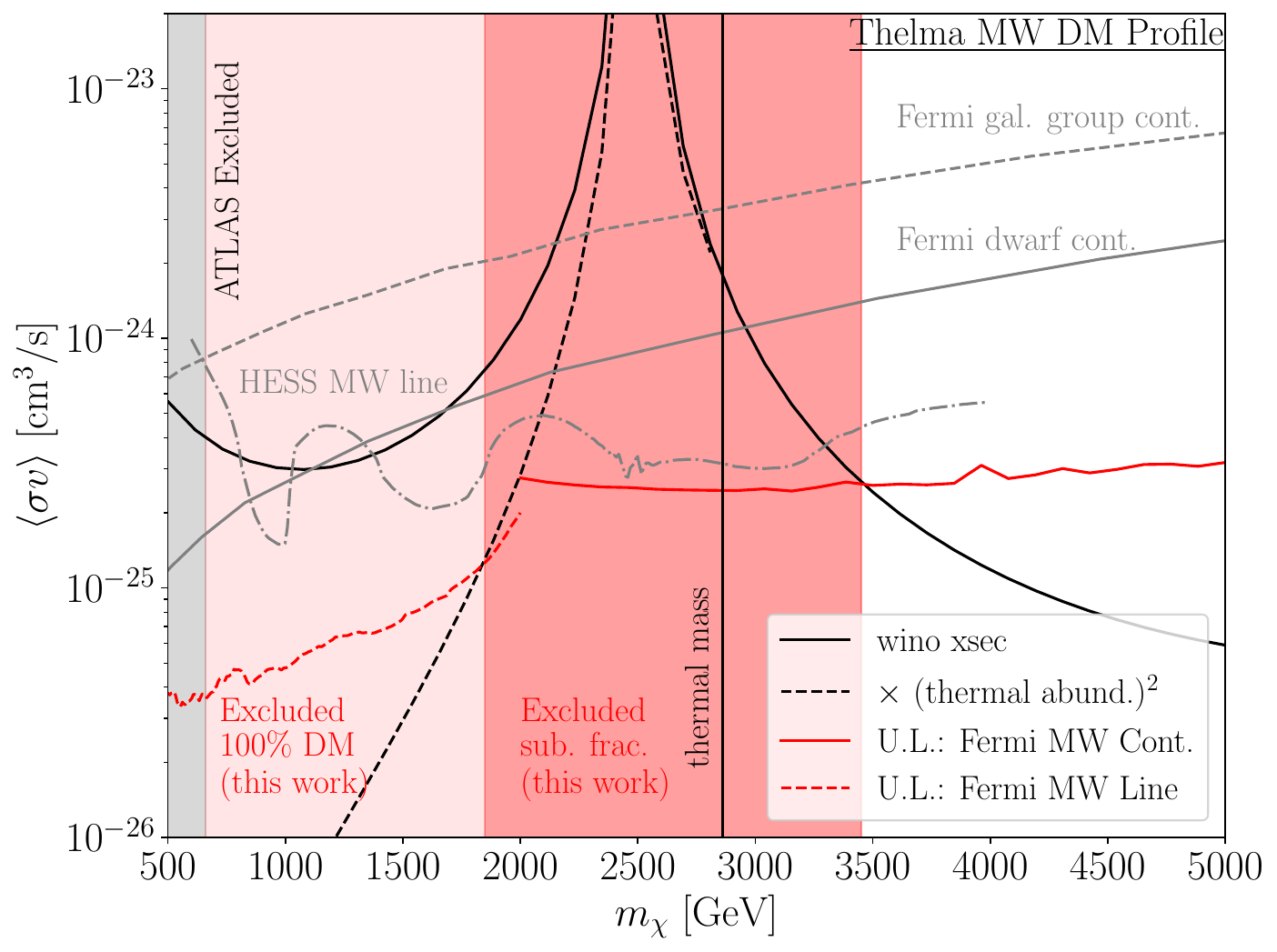}
\includegraphics[width=0.49\linewidth]{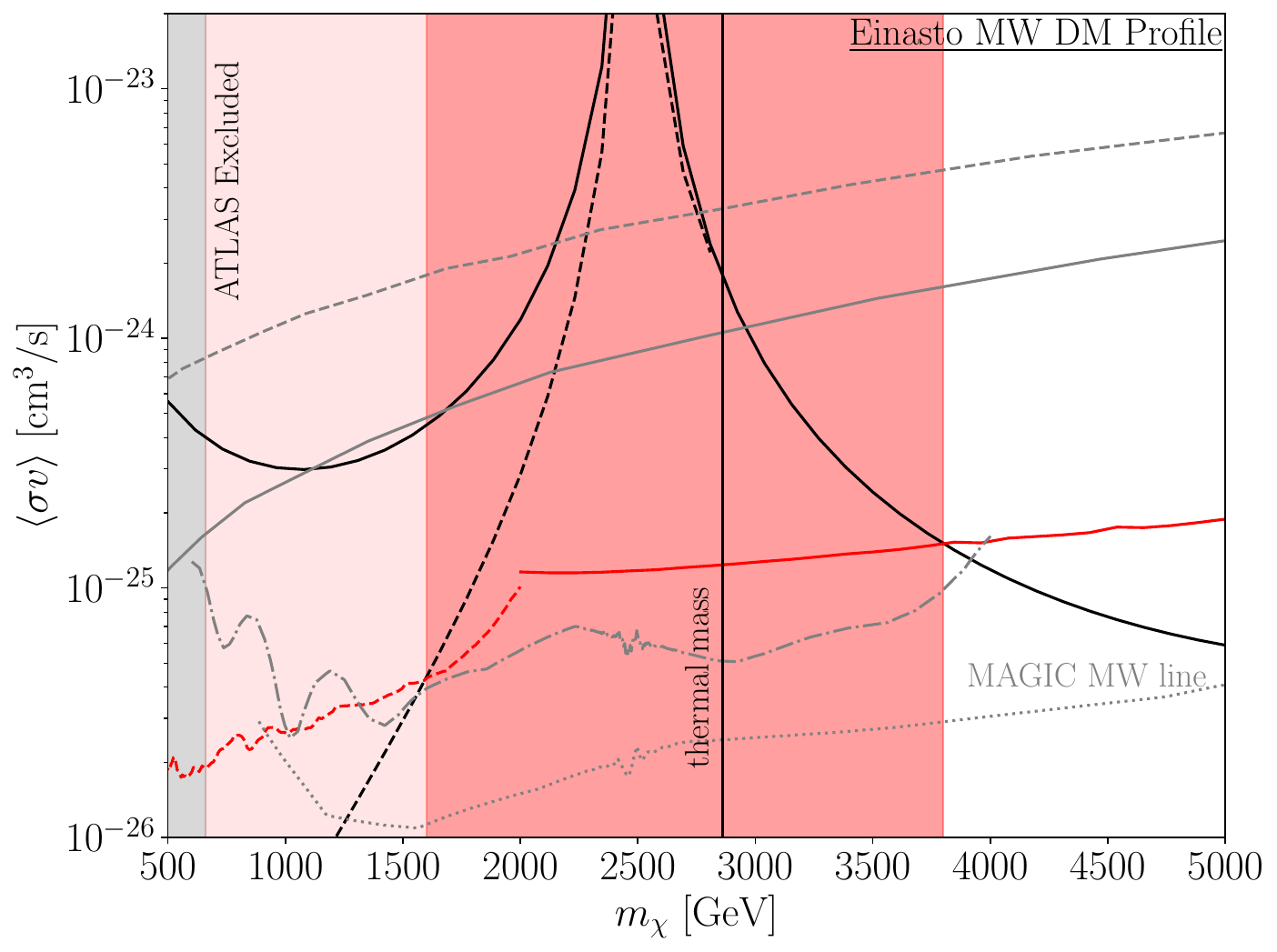}
\caption{
(Left) The 95\% upper limit on the total wino annihilation cross-section (red), assuming the conservative Thelma DM density profile for the Milky Way, as a function of the wino mass from the dedicated search for continuum gamma-rays in the inner Galaxy above 30 GeV with Fermi gamma-ray data performed in this work. Below 2 TeV we search instead for the narrow spectral endpoint at $E \approx m_\chi$ by re-interpreting the results of the search in Ref.~\cite{Foster:2022nva} in the context of the Thelma DM density profile; we illustrate these upper limits in the context of the total annihilation cross-section.  The total wino annihilation cross-section as a function of mass is shown in solid black, with the dashed black curve additionally showing the suppression $f_\chi^2$, with $f_\chi$ the expected sub-fraction of wino DM for $m_\chi < 2.86$ TeV under the standard cosmological history pre-BBN. We rule out the dark shaded (lightly shaded) mass range assuming the wino is a thermal sub-fraction (100\%) of the DM. Additionally, we show 95\% upper limits on the annihilation cross-section from the search for gamma-ray lines at the Galactic Center with HESS data~\cite{Rodd:2024qsi}, continuum emission from nearby dwarf galaxies with Fermi data~\cite{Fermi-LAT:2015att}, and continuum emission from nearby galaxy groups with Fermi data~\cite{Lisanti:2017qlb}; the former two searches each independently rule out thermal wino DM. (Right) As in the left panel but assuming the Einasto DM profile for the Milky Way; note that the galaxy groups and Fermi dwarf searches are unaffected.  In this case we also show the upper limits from the ground-based MAGIC telescope~\cite{MAGIC:2022acl}, which searched for gamma-ray lines like HESS. We stress here that all ``Line'' constraints (Fermi, HESS, and MAGIC) have been reinterpreted to appear on this plot in a model-specific way, using the expected relationship between the line and continuum cross-sections particular to the wino. We are unable to show the MAGIC results for the Thelma DM profile because the work~\cite{MAGIC:2022acl} does not provide sufficient details to do that reanalysis. 
}
\label{fig:wino_money_0}
\end{figure*}

We find no evidence for continuum emission associated with wino DM and therefore set upper-limits on $\langle \sigma v \rangle$ as a function of the mass $m_\chi$, with results illustrated in Fig.~\ref{fig:wino_money_0}.  In the left panel we show our results (red, solid) assuming the Thelma DM profile, which is the most conservative from the ensemble of FIRE-2 Milky Way analogue galaxies we consider.  Even with Thelma, we rule out the thermal wino by around an order of magnitude in the cross-section. Note that we show the total (tree-level with  Sommerfeld enhancement) wino annihilation cross section as a function of mass in solid black.  In the right panel we illustrate our results assuming the Einasto DM profile, which strengthens the limit at the thermal mass by approximately a factor of two.  Note that our results are robust to modest changes to the analysis framework, some of which are discussed more below. For example, if we restrict to energies between $50 \, \, {\rm GeV} < E < 500 \, \, {\rm GeV}$ and to radii $r < 10^\circ$ from the Galactic Center, we exclude the $m_\chi = 2.86$ TeV wino assuming the Thelma DM profile with $\mu < 0.9$  at 95\% CL, with ${\rm sign}(\hat \mu) \times {\rm TS} \approx 0.002$ consistent with zero evidence for DM.  

As an alternate analysis strategy to test the dependence on our assumed Galactic diffuse emission model, we use a data-driven approach whereby we include the annuli in the inner $10^\circ$ in the analysis, with the background model given by the stacked data between $20^\circ$ and $30^\circ$. (Note that we use a single spectral template for the background model, though the background-model normalization is allowed to float independently in each annulus.)  We lower the low-energy threshold to 10 GeV, as in~\cite{Dessert:2022evk}, to maintain the target sensitivity with the reduced ROI and to account for the sensitivity loss in accounting for the fact that the background model has a component of the signal.  We find no evidence for DM for any of the DM profiles, with ${\rm sign}(\hat \mu) \times {\rm TS} \approx 2$ with $\mu < 0.4$ at 95\% confidence for the Thelma DM profile.

\begin{figure}[!tb]
\centering
\includegraphics[width=1\linewidth]{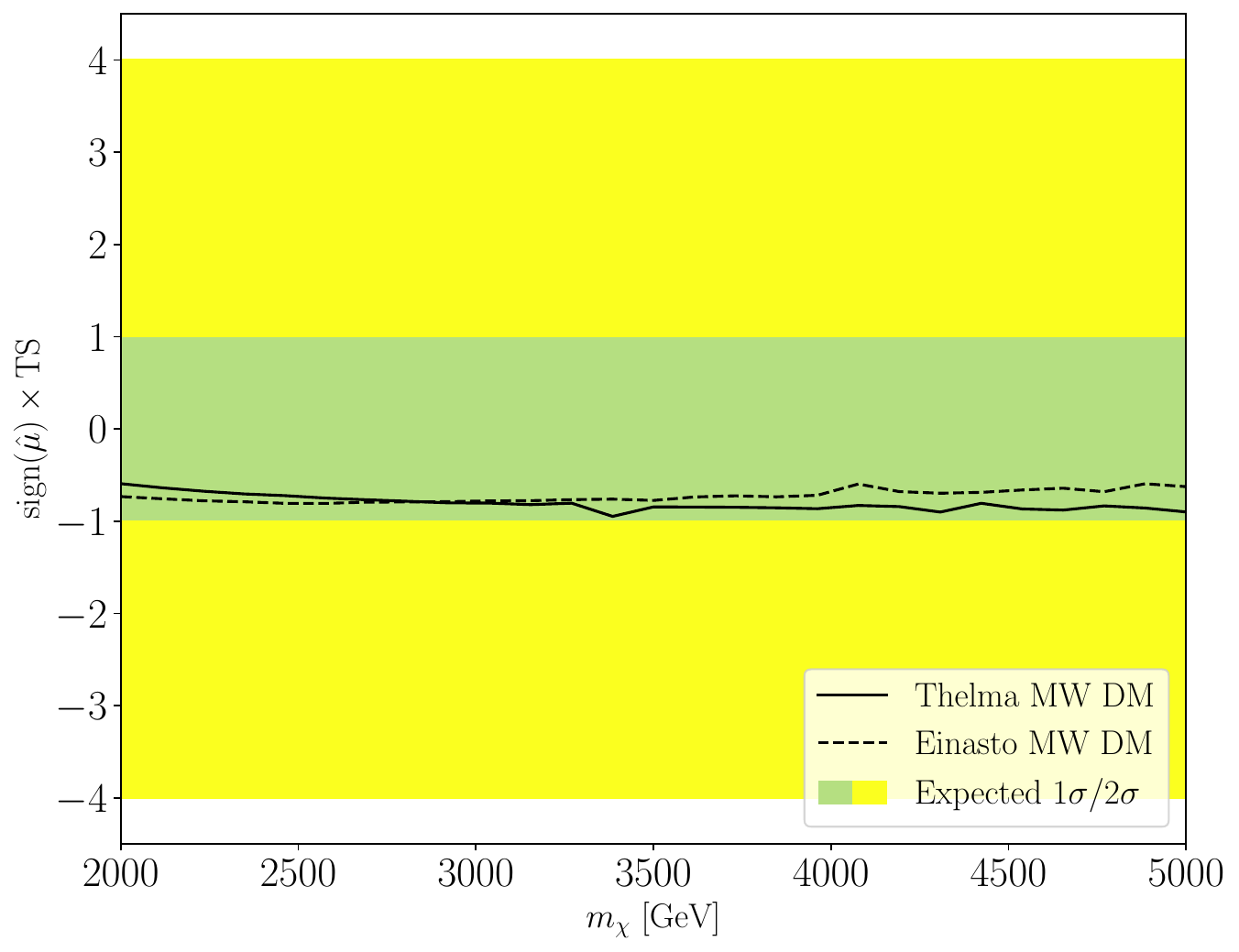}
\vspace{-0.7cm}
\caption{The discovery TS for the two-sided test in favor of the wino model as a function of mass $m_\chi$ for our fiducial analysis searching for continuum emission from wino annihilation in the inner Galaxy. We multiply the TS by the sign of the best-fit cross-section. For both the Thelma and Einasto DM profiles we find no evidence in favor of DM, with slightly negative best-fit cross-sections consistent with zero at less than 1$\sigma$ significance. }
\vspace{-0.5cm}
\label{fig:TS_wino}
\end{figure}

To emphasize the point that we find no evidence in favor of DM, in Fig.~\ref{fig:TS_wino} we show the discovery TS times the sign of the best-fit cross-section as a function of the wino mass assuming both the Thelma and Einasto DM profiles for our fiducial analysis. Both DM profiles return similar results, with the best-fit $\hat\mu$ negative across the masses probed but at less than 1$\sigma$ significance.  In principle, we power-constrain our limits~\cite{Cowan:2011an}, whereby we prevent the upper limit from fluctuating below the lower expected 1$\sigma$ expectation for the limit, though in practice since ${\rm TS} < 1$ for the two-sided test, the limits are not affected by power-constraining.

As an aside, we note that in contrast to our results  Ref.~\cite{Dessert:2022evk} found $\sim$2$\sigma$ evidence for continuum annihilation from thermal higgsino DM annihilation in nearly the same data set as used here to search for the wino.  However, there are crucial differences between the search in~\cite{Dessert:2022evk} and that here which explain the different results. First, the thermal higgsino is significantly lower in mass than the wino, with a mass $\sim$1.1 TeV, resulting in the continuum spectrum peaking at much lower energies. Secondly, because of the softer spectrum, Ref.~\cite{Dessert:2022evk} had a lower low-energy threshold than that here of $\sim$10 GeV.  Repeating our analysis with a $\sim$10 GeV low-energy threshold and the identical spatial ROI as in Ref.~\cite{Dessert:2022evk}, consisting of all annuli out to $10^\circ$, we find a positive best-fit cross-section with ${\rm TS} \approx 1.6$ in favor of the thermal wino for the Thelma DM profile, which is slightly lower than the evidence found in~\cite{Dessert:2022evk} for the higgsino due to the different spectra. However, the magnitude of that best-fit is much smaller than the wino prediction, and the 95\% upper limit under the Thelma DM profile is nearly the same as that found in our fiducial analysis, with $\mu^{95}_{\rm Thelma} \approx 0.14$ (compared to $\mu^{95}_{\rm Thelma} \approx 0.12$ in our fiducial analysis, see Tab.~\ref{tab:MDM}).   

In dashed black in Fig.~\ref{fig:wino_money_0} we show the annihilation cross-section multiplied by the expected sub-fraction squared of DM for $m_\chi$ less than the thermal mass. In particular, for a given $m_\chi < 2.86$ TeV we may calculate, under the standard cosmology, the sub-fraction of DM $f_\chi$ that the wino comprises; to good approximation, $f_\chi \approx (2.86 \, \, {\rm TeV} / m_\chi)^2$. We then show $f_\chi^2 \times \langle \sigma v \rangle$, which falls off rapidly at lower wino masses given that the wino quickly becomes a negligible sub-fraction of DM.  For example, under a thermal cosmology at $m_\chi = 1$ TeV the wino is only expected to comprise $f_\chi \approx 0.1$ of the DM, which suppresses the annihilation signal by $f_\chi^2 \approx 10^{-2}$ relative to the case where the wino comprises all of the DM.

For $m_\chi < 2$ TeV, stronger upper limits on the wino with Fermi data are instead obtained by searching for the spectral endpoint, that is, the hard line-like signal at energies $E \approx m_\chi$.  We consider the endpoint only for $m_\chi < 2$ TeV in the context of Fermi, because the Fermi instrument response does not reliably extend to higher energies. We make use of the results computed in~\cite{Foster:2022nva}, who searched for gamma-ray lines from DM annihilation in the inner Milky Way. In particular, Ref.~\cite{Foster:2022nva} used a similar ROI and data set to that in this work, although they subdivided the data into finer energy bins (in particular, 531 logarithmically-spaced energy bins between 10 GeV and 2 TeV).  The data were also subdivided into annuli of width $1^\circ$ from zero to 30$^\circ$, with the Galactic plane masked. The data were analyzed separately in each of the top three quartiles of data, as quantified by the energy resolution (\texttt{edisp}), and in each annulus; the results from these individual analyses were joined in the context of a joint likelihood. For a given DM mass, the search in a given data set consisted of a search with a Poisson likelihood for gamma-ray lines at $E = m_\chi$, forward modeled through the finite detector energy resolution response, with the background described by a power-law model in a narrow energy range centered around $m_\chi$. (Note that the power-law background models were given independent nuisance parameters when analyzing each data set.) At high energy ($m_\chi \gtrsim 500$ GeV) that work found that the analysis was effectively in the zero-background-counts limit.  Ref.~\cite{Foster:2022nva} presents their results in terms of upper limits on $\langle \sigma v \rangle_{\gamma\gamma}$---the annihilation cross-section to two photons---as a function of $m_\chi$.  However, that work only presents results assuming an NFW DM profile and not the Thelma DM profile or the Einasto DM profile.  We reprocess the results from the individual annuli in Ref.~\cite{Foster:2022nva} to weight the joint likelihood over all annuli assuming different DM profiles, in particular the Thelma and Einasto profiles.  

Our reprocessed Fermi gamma-ray line search results are shown in Fig.~\ref{fig:wino_money_0} assuming the Thelma DM profile (left) and Einasto profile (right). 
 While these limits are computed in terms of the partial line-like annihilation cross-section to two photons ($\langle \sigma v \rangle_{\gamma\gamma}$), we present the results in terms of the total annihilation cross-section $\langle \sigma v \rangle$.  We are able to do this because for a given $m_\chi$ we may compute both the total annihilation cross-section and the annihilation cross-section to two-photons $\langle \sigma v \rangle_{\gamma\gamma}$ accounting for Sommerfeld enhancement and also end-point contributions (see Sec.~\ref{sec:min} for details). Under the assumption that the wino constitutes 100\% of the DM for $m_\chi < 2.86$ TeV, in which case non-thermal production mechanisms are required~\cite{Cohen:2013ama}, we exclude all wino masses down to the ATLAS upper limit of $\sim$660 TeV~\cite{ATLAS:2022rme}, which is also indicated in the Figure. This rescaling of limits is repeated for the HESS and MAGIC constraints, which are similarly focused on line-like signals instead of the full continuum. To emphasize the point, while the other gamma-ray constraints in Fig.~\ref{fig:wino_money_0} are roughly similar to a search for a model-independent $W^+W^-$ signal, the relative placement of any `Line' constraints on the figure is very much wino-specific. 

The total excluded mass range for the wino in our analysis, assuming the wino constitutes 100\% of the DM, is shaded in light red; we exclude all wino masses below $m_\chi \lesssim 3.45$ TeV for the more conservative Thelma DM profile.  In the case where the wino constitutes a sub-fraction of the DM, with the sub-fraction $f_\chi$ determined under the standard radiation-dominated cosmological history pre-BBN, then we are unable to exclude wino masses above the ATLAS bound for $m_\chi \lesssim 1.85$ TeV, assuming the Thelma DM profile. (Note that in these scenarios the DM sub-fraction is less than 50\%.)  Winos in this mass range are detectable by future colliders, such as a future muon collider or the 100 TeV FCC-hh~\cite{Han:2020uak,Capdevilla:2021fmj}, which are entirely insensitive to whether the wino makes up any part of the DM.   

It is worth focusing specifically on the thermal wino ($m_\chi \approx 2.86$ TeV) to address the question of how robustly this particular model is excluded. Towards that end, in Fig.~\ref{fig:wino_money_0} we reinterpret the HESS gamma-ray line search at the Galactic Center from~\cite{Rodd:2024qsi} in terms of the Thelma DM profile (left panel) and Einasto DM profile (right panel). We caution, however, that the result from Ref.~\cite{Rodd:2024qsi} is {\it weaker} than the published result from the HESS collaboration~\cite{HESS:2018cbt}, which Ref.~\cite{Rodd:2024qsi} calls into question due to a possible error in the HESS analysis pipeline affecting the results of a later collaboration work~\cite{HESS:2022ygk}.  Ref.~\cite{Rodd:2024qsi} analyzed 546 hr of HESS data from the inner Galaxy that was made available in~\cite{HESS:2022ygk}; in contrast, the HESS collaboration gamma-ray line search in~\cite{HESS:2018cbt} used an older data set of only half the exposure time, making a stronger constraint difficult to understand. Note however that Ref.~\cite{HESS:2018cbt}, unlike~\cite{HESS:2022ygk}, does not provide sufficient information for a direct reproduction of their analysis.

Among the various analyses performed in Ref.~\cite{Rodd:2024qsi}, one that focused on the $\gamma\gamma$ line signature utilized HESS data in 8 Galactocentric annuli extending from 0.5$^\circ$ to 3$^\circ$ and performed a search for a narrow gamma-ray line in each annulus within a sliding energy window, modeling the background in each annulus with an independent power law and combining the results between annuli with a joint likelihood. We present results for the Thelma and Einasto profiles, which enter into the joint likelihood via different weightings of annuli. Note that while the analyses presented in Refs.~\cite{HESS:2022ygk,HESS:2018cbt} employ an ON-OFF strategy for background subtraction -- whereby for a given HESS pointing an ON region is defined towards the Galactic Center and an OFF region is defined further away from the Galactic Center but at a symmetric location with respect to the HESS beam axis -- we use here only the ON data for our constraints. The benefit is to avoid the severe signal loss in ON-OFF subtraction analyses for cored DM scenarios, rendering our results much less sensitive to the choice of density profiles. We refer the reader to~\cite{Rodd:2024qsi} for details of this procedure. This analysis, like that performed in~\cite{Rodd:2024qsi}, returns no evidence for DM.

Given the controversy regarding the HESS results, however,
 they should be treated with some level of caution. With that said, the reprocessed HESS results from~\cite{Rodd:2024qsi} exclude the thermal wino even for the Thelma DM profile, and if the official collaboration results presented in~\cite{HESS:2018cbt} are indeed correct the wino would be even further excluded.  
Note that the HESS analysis is more sensitive to the assumed DM profile than Fermi, given that HESS is a pointing instrument that focuses more on the very inner regions of the Galaxy relative to Fermi. As alluded to above, this is exacerbated by the data-subtraction technique employed by official HESS analyses, which substantially degrades the sensitivity for cored DM profiles by subtracting away large amounts of observed signal.

The similarly ground-based Major Atmospheric Gamma-ray Imaging Cherenkov (MAGIC) telescopes have also searched for gamma-ray lines at the Galactic Center~\cite{MAGIC:2022acl}, and we include their results assuming the Einasto profile in the right panel of Fig.~\ref{fig:wino_money_0}.  The upper limits from MAGIC are even stronger than those from HESS and those from our analyses in this work. On the other hand, we are unable to reinterpret these results in the context of the Thelma DM profile because Ref.~\cite{MAGIC:2022acl} does not provide sufficient information for that reanalysis. We expect, however, that the drop in sensitivity going from Einasto to Thelma would be similar to that seen for HESS, in which case MAGIC would likely also constrain the thermal wino in the case of the Thelma profile.

In Fig.~\ref{fig:wino_money_0} we show the upper limits on the total wino annihilation cross-section from the Milky Way dwarf galaxy search presented by the Fermi Collaboration in~\cite{Fermi-LAT:2015att} for the $W^+W^-$ final state.  Note that while wino annihilation is primarily to the $W^+W^-$ final state, there is also a small contribution from the $ZZ$ channel. However, given that~\cite{Fermi-LAT:2015att} does not present results for $ZZ$ annihilation, we assume in this case that wino annihilation is reasonably approximated by a 100\% branching ratio to $W^+W^-$. This approximation is justified because the limits from the $W^+W^-$ and $ZZ$ final states tend to match each other to around 10\% accuracy, given that the two induced gamma-ray spectra are similar (see, {\it e.g.},~\cite{Hoof:2018hyn}), and given that $W^+W^-$ dominates over $ZZ$ for the wino.  The 95\% upper limits on $\langle \sigma v \rangle$ for the wino from the dwarf search are the same in both panels of Fig.~\ref{fig:wino_money_0} given that this analysis does not invoke the DM profile of the Milky Way. Importantly, the Milky Way dwarf search independently excludes the thermal wino. This is reassuring because the dwarf galaxies targeted in~\cite{Fermi-LAT:2015att} are DM-dominated systems, which are not expected to be affected by baryonic feedback, and thus the results in~\cite{Fermi-LAT:2015att} are both independent of those from the Milky Way Galactic Center and also potentially less subject to systematic uncertainties from misunderstanding the DM distribution in these systems. Additionally, the Milky Way dwarf galaxies have lower gamma-ray backgrounds than the Galactic Center and less spatial extension, simplifying the data analyses. 
 We note that the Milky Way dwarf search excludes almost all DM masses below the thermal mass assuming the wino is 100\% of the DM, except for a narrow range of masses around $\sim$1.3 TeV, though this mass range is strongly excluded by the search for gamma-ray lines in the Milky Way.

 In Fig.~\ref{fig:wino_money_0} we also show the 95\% annihilation constraints on the total wino annihilation cross-section from the Fermi galaxy group search in~\cite{Lisanti:2017qoz,Lisanti:2017qlb}, again reinterpreted from the upper limits presented in terms of $W^+W^-$. That work performed a similar analysis to the dwarf search~\cite{Fermi-LAT:2015att} discussed above, though focusing on nearby galaxy groups such as the Virgo cluster, NGC0253, NGC3031, and the Centaurus cluster, among others. That work found no evidence for DM annihilation with results show in Fig.~\ref{fig:wino_money_0}. These results are not strong enough to independently rule out the thermal wino, though they do constrain lower-mass winos. Note that Ref.~\cite{Lisanti:2017qlb} excluded M31 from their fiducial analysis because of its large spatial extent, though a subsequent analysis focusing on M31 has since found strong limits on M31 that may independently exclude the thermal wino~\cite{DiMauro:2019frs}, though unfortunately that work does not present results for the $W^+ W^-$ final state.

 \begin{figure}[!tb]
\centering
\includegraphics[width=1\linewidth]{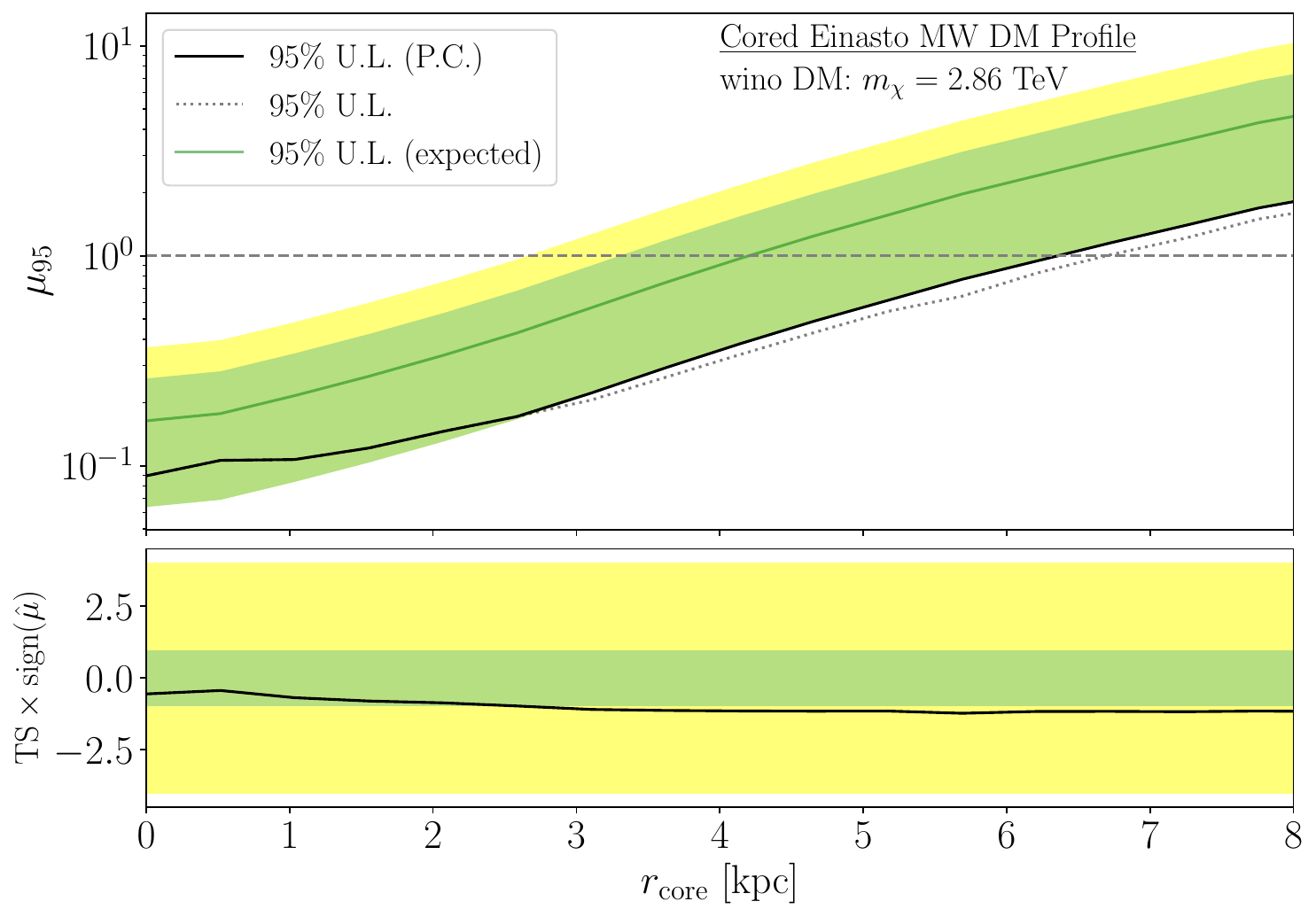}
\vspace{-0.7cm}
\caption{(Upper) The 95\% upper limit on the annihilation cross-section, reported in terms of the $\mu$ parameter that multiplies the expected wino cross-section, for the thermal wino with $m_\chi = 2.86$ TeV as a function of the core size $r_{\rm core}$ for the assumed DM density profile that follows the Einasto profile until distances $r = r_{\rm core}$ from the Galactic Center and is constant for lower distances. The green and gold bands show the expected power-constrained upper limit at 1/2$\sigma$ containment under the null hypothesis.  At large $r_{\rm core}$ our limits become power-constrained, though the un-constrained limits are shown in dotted black and are only slightly stronger than the power-constrained limits. (Bottom) As in Fig.~\ref{fig:TS_wino} but for the analysis described in the upper panel. The discovery TS becomes slightly larger than unity at core sizes larger than around 2 kpc, with a slightly negative best-fit cross-section.   }
\vspace{-0.5cm}
\label{fig:core_wino}
\end{figure}

We may ask the question of how large of a Milky Way core size would be needed in order for the thermal wino to be consistent with the Fermi data. As discussed further in Sec.~\ref{sec:DM}, here we consider an Einasto DM profile for the Galaxy with the DM profile constant within the inner radius $r_{\rm core}$. In Fig.~\ref{fig:core_wino} we repeat our fiducial analysis procedure focusing specifically on the thermal wino ($m_\chi = 2.86$ TeV) and varying the DM core size.  We show that we are able to rule out the thermal wino at 95\% confidence for all core sizes $r_{\rm core} \lesssim 6.7$ kpc. That is, we are able to rule out thermal wino DM assuming a DM core so large that it extends almost to the solar radius, in which case the DM density never rises above $\rho \approx 0.5$ GeV$/$cm$^3$.  Such a large DM core is not reproduced in any DM simulation and has no physical motivation, allowing us to conclude that the thermal wino is robustly excluded even accounting conservatively for uncertainties on the DM profile.  To emphasize this point, while in this work we assume a local DM density of 0.38 GeV/cm$^3$, even if this were lower by around a factor of two at $\rho_\odot = 0.2$ GeV/cm$^3$, which is clearly in tension with all measurements of the local DM density as discussed in the previous section, we would be able to rule out thermal wino DM for core sizes lower than $r_{\rm core} \lesssim 3.7$ kpc, which is also unphysically large.  

Note that, as shown in the bottom panel of Fig.~\ref{fig:core_wino}, the TS in favor of the wino model becomes slightly above 1, with a negative best-fit cross-section, at large core sizes. As a result, the limit becomes power constrained, although only slightly, as seen in the upper panel, where we show both the power-constrained and original limits. 

\begin{figure}[!htb]
\centering
\includegraphics[width=1\linewidth]{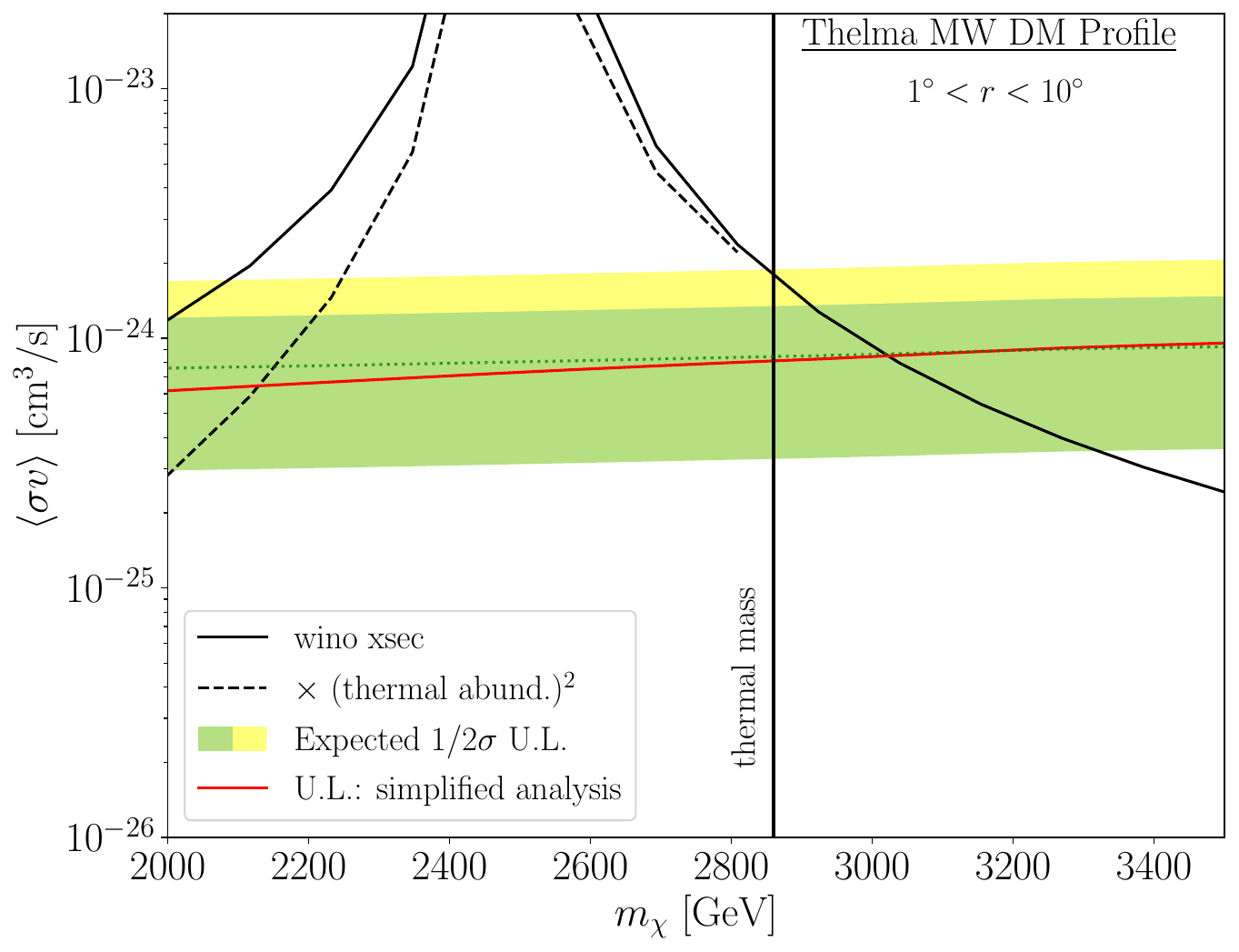}
\vspace{-0.7cm}
\caption{The 95\% upper limit (red) on the wino annihilation cross-section, as in Fig.~\ref{fig:wino_money_0}, but for our alternate analysis whereby we analyze a single, stacked ROI that consists of all unmasked data with $r < 10^\circ$ from the Galactic Center.  The data for this analysis are illustrated in Fig.~\ref{fig:Fermi_data}. 
 We find no evidence for DM from this analysis, which is less sensitive than our fiducial analysis, with a discovery TS near zero. Still, this analysis rules out the thermal wino, serving as a cross-check of our results.  }
\vspace{-0.5cm}
\label{fig:simplified}
\end{figure}

Lastly, to emphasize the robustness of our results for the thermal wino, we repeat our wino analysis with the simplified analysis framework described in Sec.~\ref{sec:data} whereby we stack the spectral data over the full ROI of the inner 10$^\circ$, subject to our fiducial plane and PS masks, and perform a single spectral search over this region for wino DM.  The upper limit from this analysis is shown in Fig.~\ref{fig:simplified} assuming the Thelma DM profile.  The simplified analysis, which is less sensitive than our fiducial analysis given the lack of spatial information, still strongly excludes the thermal wino. Moreover, at the thermal wino mass the discovery significance in favor of the wino model for the two-sided test is nearly zero, with a best-fit cross-section nearly identically zero.  

\section{Results for minimal DM with $n \geq 5$}
\label{sec:nb5}

\begin{figure}[!tb]
\centering
\includegraphics[width=1\linewidth]{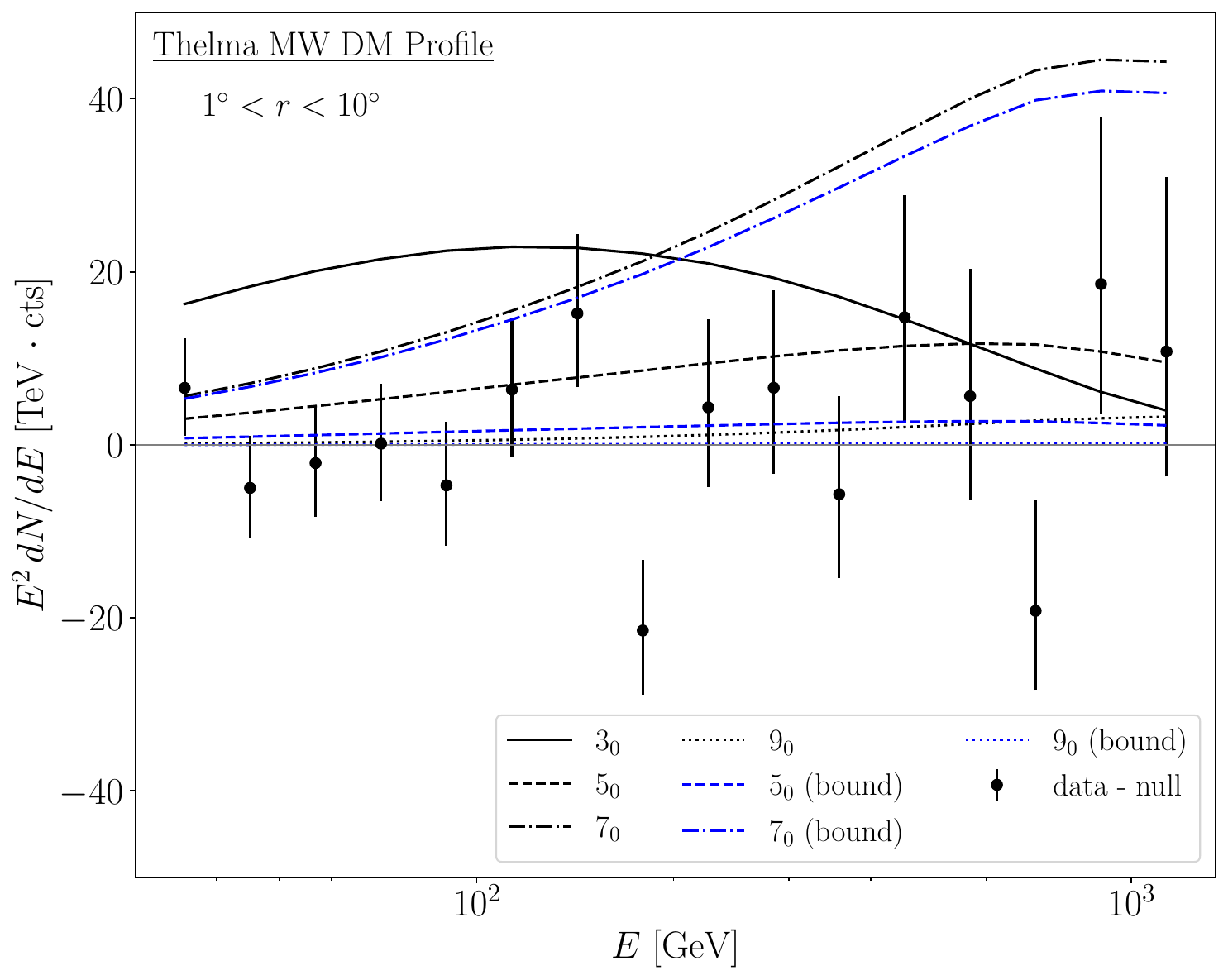}
\vspace{-0.7cm}
\caption{As in the lower panel of Fig.~\ref{fig:Fermi_data} but comparing all of the minimal DM models we consider to the background-subtracted data. (Note that the background subtracted data in the inner $10^\circ$ are shown for illustrative purposes only.) At higher $n$, the minimal DM models have continuum spectra that are peaked at higher energies. Here, we show the predictions at the central mass values given in Tab.~\ref{tab:MDM}, though it is important to account for uncertainties in the thermal masses because of the presence of Sommerfeld resonances across this allowed mass range.  Note that we show separately the contributions from bound-state formation and decay in blue, while the total continuum spectra are illustrated in black. (See text for details.) }
\vspace{-0.5cm}
\label{fig:bkg_sub_multi}
\end{figure}

We may search for minimal DM through the same framework as we use for the wino continuum search with Fermi data. In particular, in Fig.~\ref{fig:bkg_sub_multi} we illustrate the spectra for the continuum gamma-rays in the energy range of interest for Fermi for the $3_0$, $5_0$, $7_0$, and $9_0$ DM models.  We illustrate these spectra assuming the central thermal mass values listed in Tab.~\ref{tab:MDM}, though we note that for the $n \geq 5$ models the uncertainties on the mass play an important role, as we discuss further below.  

In Fig.~\ref{fig:bkg_sub_multi} we show both the total annihilation spectra (black) along with the contributions from bound state formation and decay (blue). The relative importance of bound state formation to the spectrum is highly sensitive to the positioning of the thermal mass relative to resonances.  
For example, the central thermal mass point of the $7_0$ model is dominated by the bound state contribution, but that is not true for the entire range of allowed masses. 
This is why it is important to profile over the thermal mass uncertainties when making statements about the $n \geq 5$ models, which have much larger uncertainties than the wino for reasons discussed previously. Note, also, in Fig.~\ref{fig:bkg_sub_multi} that there is a clear trend whereby the higher-dimensional multiplets have spectra that are peaked at increasingly high energies. This is unsurprising given that the thermal masses increase with increasing $n$.

\begin{figure}[!tb]
\centering
\includegraphics[width=1\linewidth]{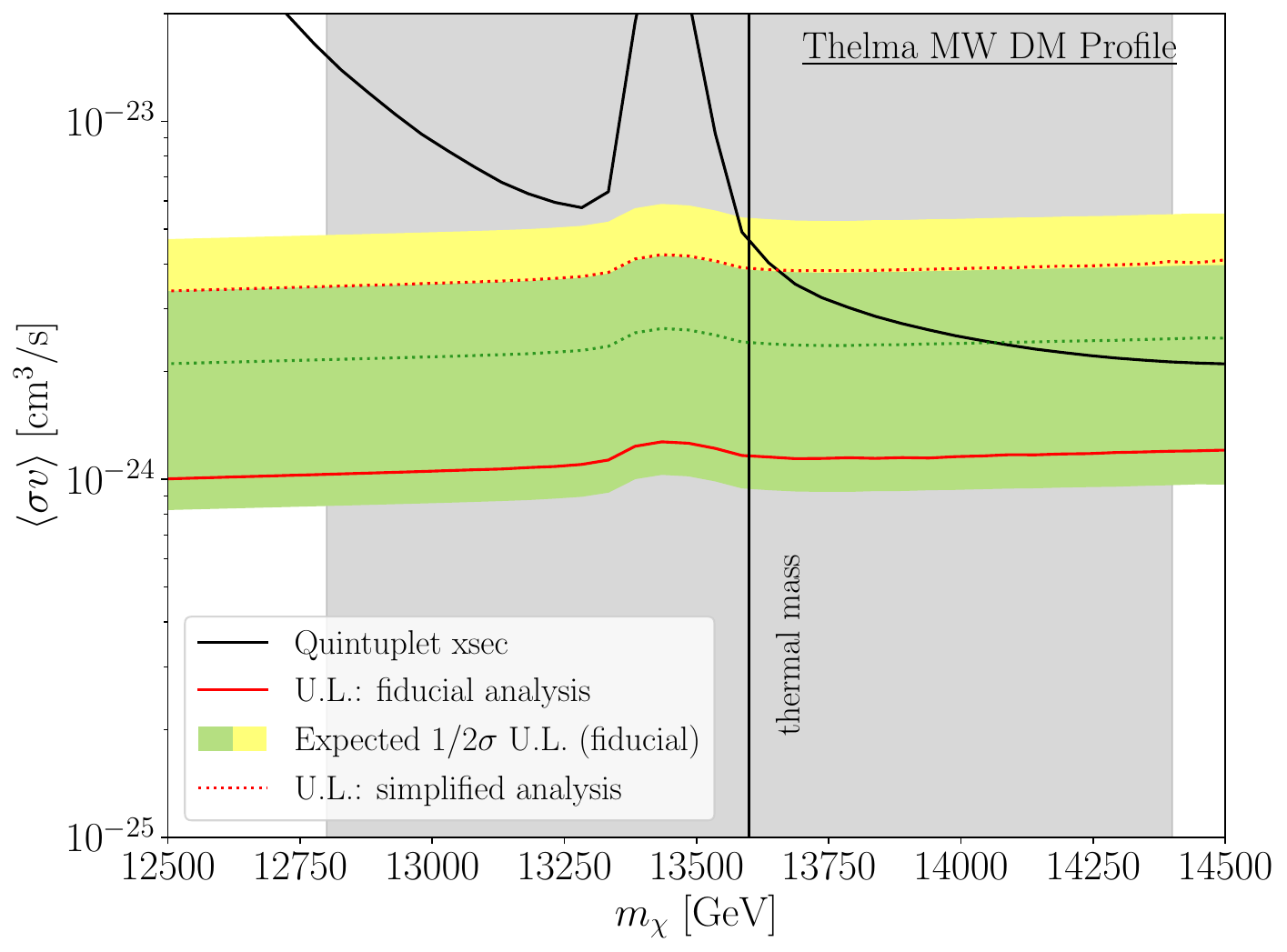}
\vspace{-0.7cm}
\caption{The upper limit (red) as a function of mass on the quintuplet ($5_0$) annihilation cross-section as a function of the quintuplet mass in the vicinity of the expected thermal mass assuming the conservative Thelma DM profile. The mass where the quintuplet can explain 100\% of the DM is expected to lie within the gray band (see Tab.~\ref{tab:MDM}). The upper limit from this analysis is compared to the quintuplet cross-section in black. We conclude that the quintuplet model is strongly disfavored by our analysis. As a cross-check, we perform the simplified analysis on the stacked inner $10^\circ$ data set, with the resulting upper limit shown in dotted red (see text for details).  }
\vspace{-0.5cm}
\label{fig:quint_limit}
\end{figure}

\begin{figure*}[!htb]
\centering
\includegraphics[width=0.49\linewidth]{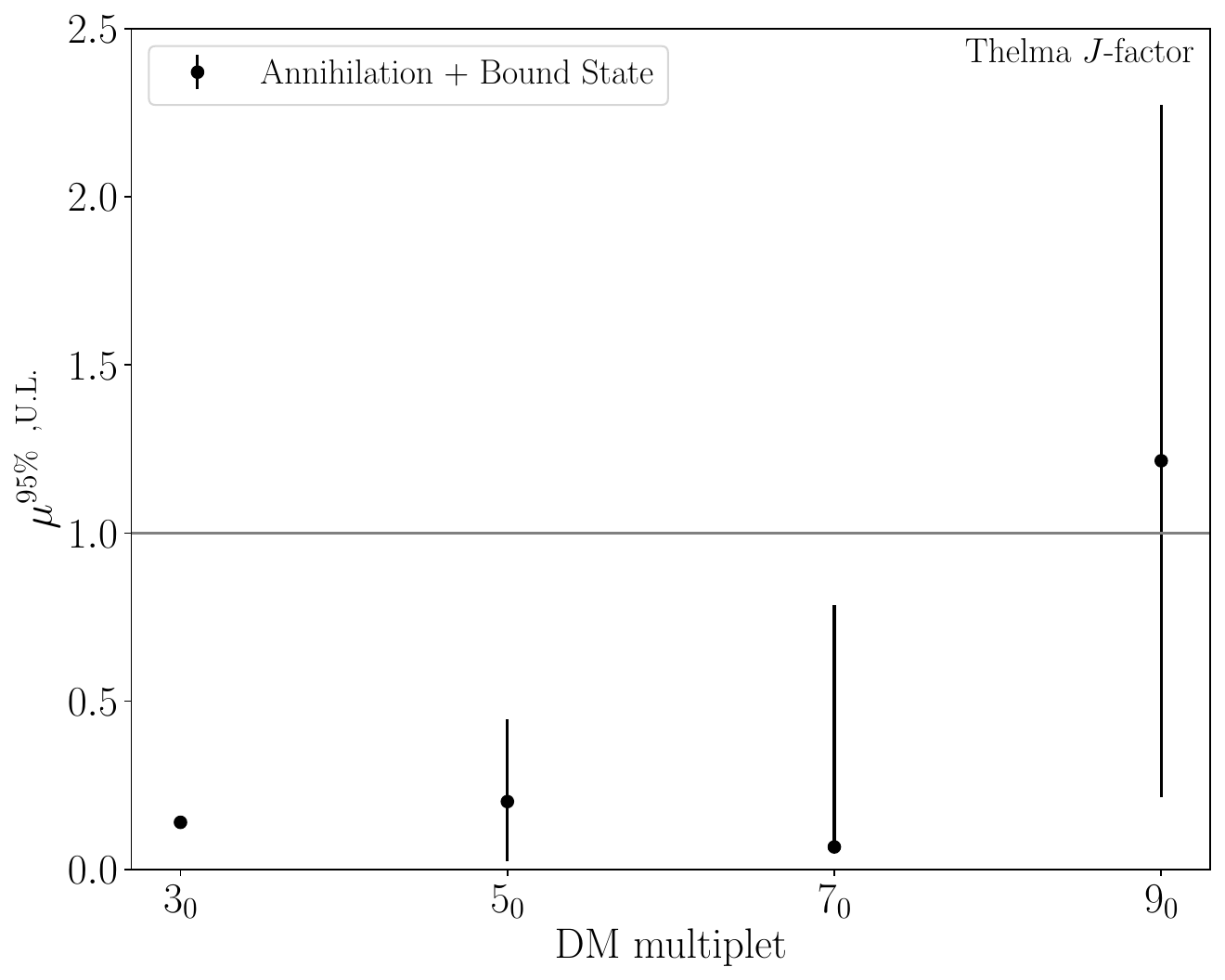}
\includegraphics[width=0.49\linewidth]{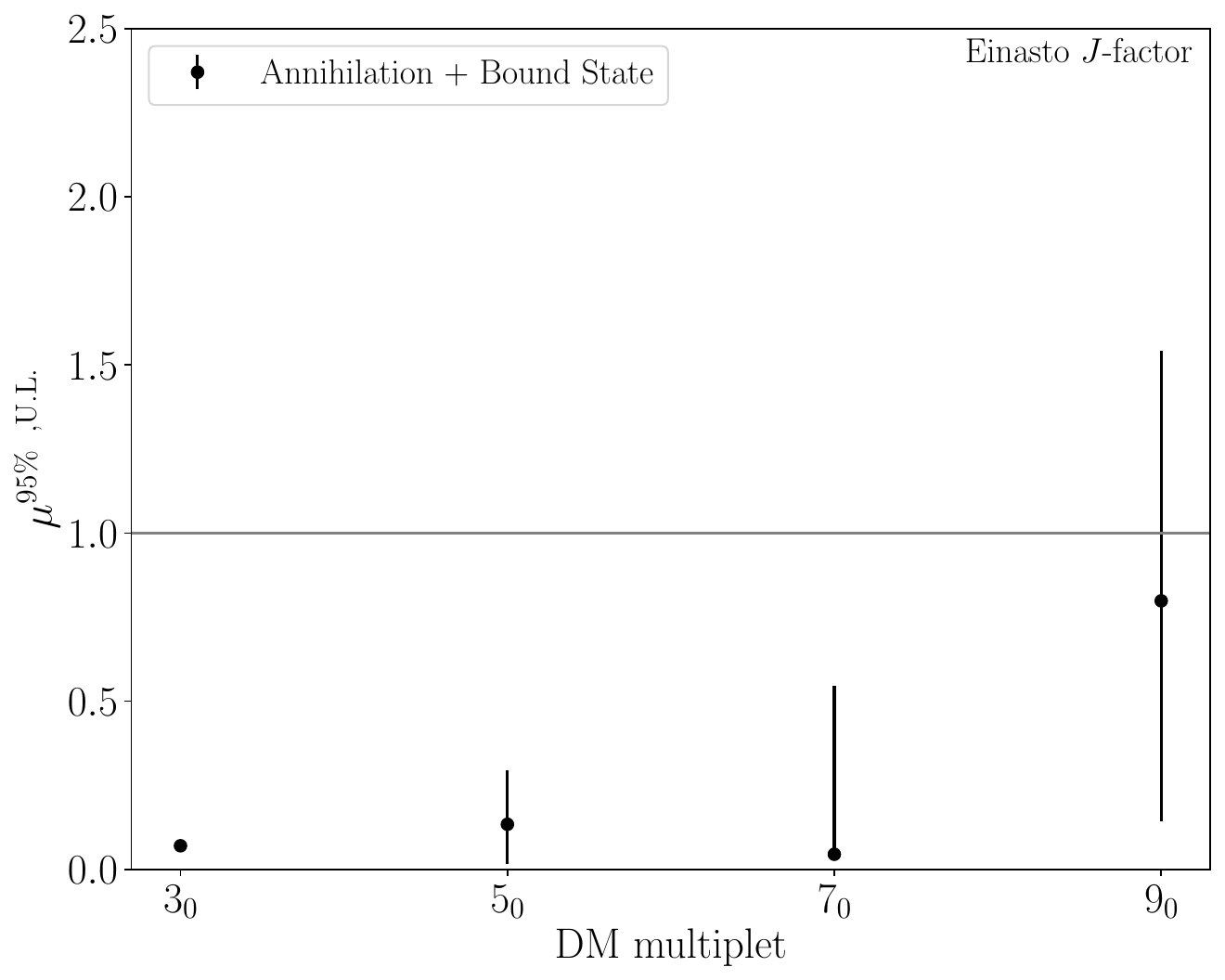}
\caption{The 95\% upper limits on the minimal DM annihilation cross-sections from our fiducial Fermi continuum data analyses assuming the Thelma DM density profile (left) and Einasto profile (right) for the Milky Way. The cross-sections are represented through the $\mu$ parameter, which rescales the expected cross-sections for the respective models. The central points correspond to central mass points given in Tab.~\ref{tab:MDM} for the thermal models that explain 100\% of the DM; the error bars arise from scanning over the allowed mass ranges quoted in Tab.~\ref{tab:MDM}. For the wino ($3_0$), quintuplet ($5_0$), and $7_0$ models we exclude the thermal masses even for our most conservative Galactic DM density profile. We thus conclude that these models are disfavored from explaining the DM in our Universe.  On the other hand, the $9_0$ model is still allowed by our analysis. }
\label{fig:money}
\end{figure*}

In Fig.~\ref{fig:quint_limit} we illustrate 
the upper limits on the total annihilation cross-section from our continuum search with Fermi data in the context of the quintuplet ($5_0$) model and assuming the Thelma DM profile. The upper limit is shown in solid red, with the expected upper limit at 1/2$\sigma$ in green and gold, respectively. In gray we shade the mass range where, at 1$\sigma$, the quintuplet is expected to comprise 100\% of the DM, with the central thermal mass estimate denoted by the vertical black line. The discovery TS in favor of the quintuplet model for the two-sided test is less than unity, with a slightly negative best-fit cross-section, implying no evidence in favor of quintuplet DM over the full mass range of interest.  Given that we rule out the quintuplet over its full possible mass range for our more conservative DM profile, we conclude that the quintuplet model is strongly disfavored as a model that explains 100\% of the DM.  To emphasize this point, we note that repeating our analysis with the cored Einasto profile we find that the quintuplet is disfavored for all core sizes $r_{\rm core} \lesssim 2$ kpc.  On the other hand, the upcoming CTAO may have discovery capability for the thermal quintuplet in the case where the Milky Way DM profile is cored at a level even beyond 2 kpc~\cite{Baumgart:2023pwn}, or if the quintuplet constitutes some sub-fraction of the total DM abundance.

As a cross-check of this analysis, we also perform our simplified analysis that uses a single ROI stacked over the inner 10$^\circ$, subject to our standard plane and PS masks. (Note that the best-fit null-model-subtracted data are compared to the quintuplet spectrum in Fig.~\ref{fig:bkg_sub_multi}.)  As in the case of the wino search, the simplified analysis is less sensitive than our fiducial analysis because it does not incorporate spatial information. Still, the simplified analysis returns no evidence in favor of DM (${\rm TS} < 1$), with a slightly positive best-fit cross-section. The upper limits as a function of mass from this analysis assuming the Thelma DM profile is shown in Fig.~\ref{fig:quint_limit} in dotted red. That analysis rules out the quintuplet at the central mass expectation and lower masses, though it alone is not sensitive enough to rule out the higher-mass part of the thermal mass uncertainty band. Still, it serves as a cross-check of our fiducial analysis.

\begin{figure}[!tb]
\centering
\includegraphics[width=1\linewidth]{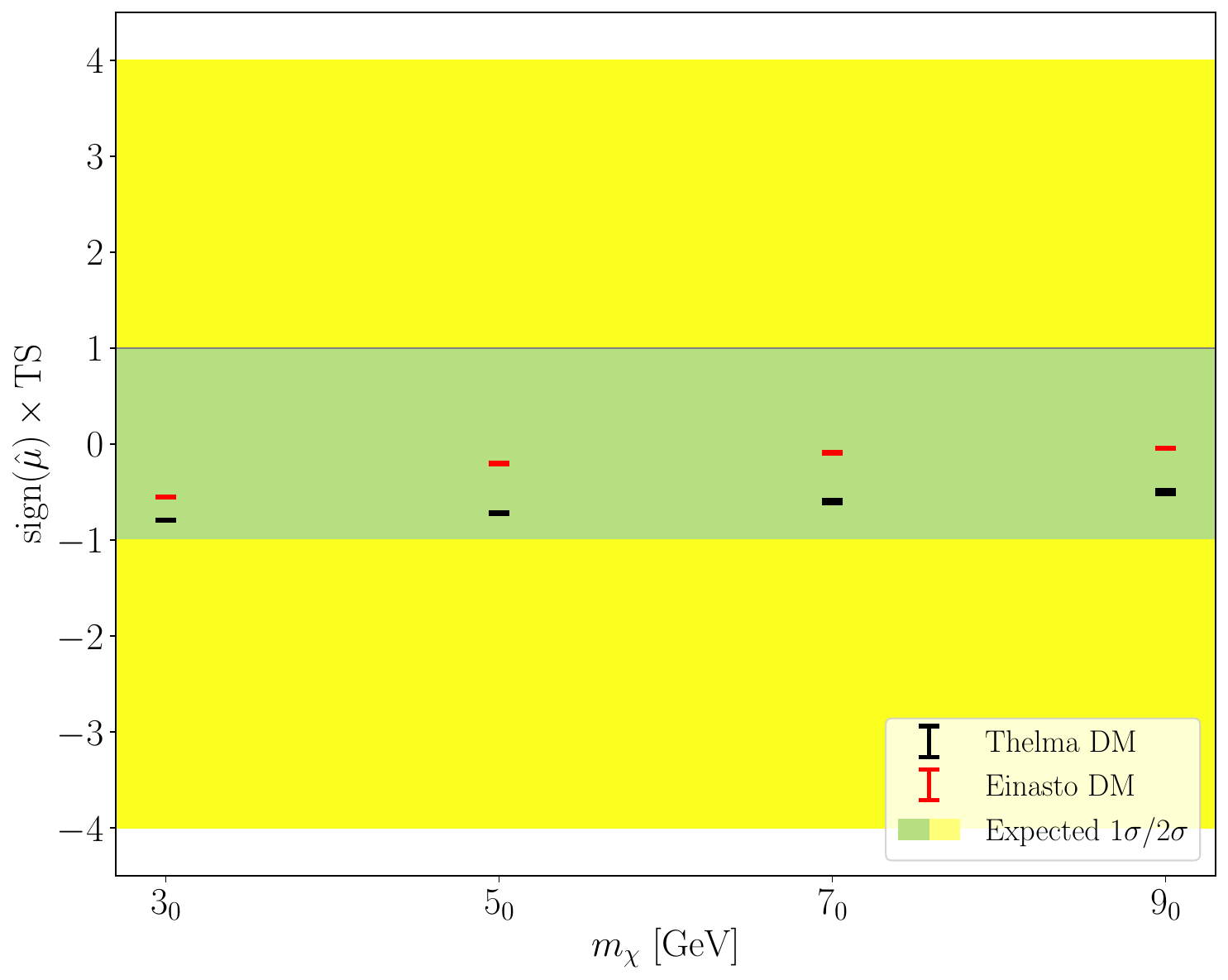}
\vspace{-0.7cm}
\caption{As in Fig.~\ref{fig:TS_wino} but for the different minimal DM models for our fiducial Fermi continuum data analyses (see Fig.~\ref{fig:money}).  We show results assuming the Thelma and Einasto Galactic DM density profiles. For a given model we show by vertical error bars the range of discovery TSs found by varying over the thermal masses in Tab.~\ref{tab:MDM}, though for most models the resulting ranges in TS are too small to be perceptible. In no case do we find evidence for DM or mismodeling. }
\vspace{-0.5cm}
\label{fig:TS_minimal}
\end{figure}

Having argued that the quintuplet model is also disfavored by our continuum analysis, we repeat the analysis strategy for the $7_0$ and $9_0$ models. The results from these searches are shown in Fig.~\ref{fig:money} and Fig.~\ref{fig:TS_minimal}.  In Fig.~\ref{fig:money} we illustrate the upper limits from our fiducial analyses applied to the $3_0$ through $9_0$ models assuming the Thelma DM profile (left) and Einasto profile (right). We present results in terms of the $\mu$ parameter, which rescales the expected minimal DM annihilation cross-section. The central points in these figures show the result for the central mass point in the uncertainty range for the expected thermal mass, with the vertical error bands showing the range of upper limits found while scanning over the full, allowed thermal mass ranges for the minimal models at 1$\sigma$. For example, this figure makes it clear that even accounting for uncertainties on the thermal quintuplet mass and in the DM profile of the Milky Way, we rule out the quintuplet model by over a factor of two in the cross-section.  In fact, we also rule out the $7_0$ model as comprising 100\% of the DM. The $9_0$ model is partially constrained by our analysis, though our analysis is not sensitive enough to fully rule out that scenario. Thus, we conclude that real, thermal, minimal DM with $n < 9$ is excluded, while in principle $n \geq 9$ models are still allowed, although we argue that they may be theoretically less appealing than the lower-$n$ scenarios.

As $n$ increases, our analyses probe increasingly high energies in the Fermi data. As such we should expect the analyses between different $n$ to be correlated but not identical. This may be seen in Fig.~\ref{fig:TS_minimal}, which shows the discovery TS for the two-sided test multiplied by the sign of the best-fit annihilation cross-section for the different minimal model analyses. We illustrate the results for both the Thelma and Einasto DM profiles. In both cases we find ${\rm TS} < 0$ for all DM models, with the TS approaching zero at higher $n$. Note that we show with vertical error bands the range of discovery TSs achieved while scanning over the allowed thermal mass ranges, though in practice these changes to the discovery TS are too small to be easily visible. We conclude that there is no evidence for mismodeling or for minimal DM in any of our analyses.

\section{Discussion}
\label{sec:discussion}

In this work we search for evidence of real, minimal DM ($n_0$ representations of $SU(2)_L \times U(1)_Y$) with Fermi gamma-ray data. We search, in particular, for the continuum gamma-rays produced in the energy range between $30$ GeV and $2$ TeV from the decays of unstable particles produced during the prompt annihilation processes and also the formation and decays of bound states. We compute these contributions precisely, accounting for Sommerfeld enhancement. We find no evidence in favor of any of the models we consider and are able to robustly rule out real minimal DM as comprising 100\% of the DM under the standard cosmological framework for $n < 9$, even assuming very conservative density profiles.  

We investigate in further detail the wino model ($3_0$), since this model is strongly motivated as a DM candidate in the context of supersymmetric extensions of the Standard Model. Unfortunately, we find that it is inconceivable that winos comprise 100\% of the DM under the standard cosmological history, as even if the DM is exceptionally cored (at more than 3.5 kpc) and the DM density is a factor of two below the measured value (less than $0.2$ GeV/cm$^3$), the thermal wino would still be in tension with our null results. For the scenario of thermal quintuplet DM, we rule out the entire allowed thermal mass range for cored profiles extending out to 2 kpc.

Winos that comprise 100\% of the DM at lower masses below $2.86$ TeV are also excluded, though if the winos only comprise a sub-fraction of the DM, as dictated by a standard thermal cosmology, they are still allowed at lower masses, below roughly $1.8$ TeV.  On the other hand, additional entropy dilution after wino freeze-out but before BBN, which is motivated {\it e.g.} in the context of heavy moduli and axion fields, could allow for much larger wino masses above 2.86 TeV without overproducing the observed DM; these scenarios are also out of reach of our Fermi analysis for $m_\chi\gtrsim 3.5$ TeV. 

Thus, while our work represents the final word on standard, thermal wino and quintuplet dark matter, higher representations of thermal minimal fermionic DM, {\it i.e.} with $n\geq 9$ and $m_\chi \gtrsim 100$ TeV, have still yet to be experimentally challenged. In addition, the door remains open for non-standard cosmologies, DM sub-fractions, or exceptionally cored density profiles (though, as discussed, the required coring is difficult to realize for the quintuplet and essentially impossible for the wino). Fortunately, many of these scenarios may be discovered by a complementary program of future colliders and the upcoming CTAO.

\begin{acknowledgments}
We thank Pouya Asadi, Matthew Baumgart, Timothy Cohen, Joshua Foster, Yujin Park, Nicholas Rodd, and Tracy Slatyer for helpful discussions. 
BRS is supported in part by the DOE award DESC0025293, and BRS acknowledges support from the Alfred P. Sloan Foundation. 
The work of WLX was supported by the Kavli Institute for Particle Astrophysics and Cosmology, and by the U.S. Department of Energy under contract DE-AC02-76SF00515.
\end{acknowledgments}

\bibliography{refs}

\end{document}